\documentclass[accepted]{acmart}

\usepackage{array} 
\usepackage{longtable} 
\usepackage{multirow}

\usepackage{makecell}
\usepackage{float}
\usepackage{subfigure} 
\usepackage{natbib}

\usepackage{graphicx} 
\usepackage{makecell} 
\usepackage{tabularx} 
\usepackage{longtable} 
\usepackage{array} 
\usepackage{booktabs} 
\usepackage{makecell} 
\usepackage{array} 
\usepackage{amssymb} 
\usepackage{pifont} 
\AtBeginDocument{%
  }

\setcopyright{acmlicensed}
\copyrightyear{2018}
\acmYear{2018}
\acmDOI{XXXXXXX.XXXXXXX}

\acmJournal{JACM}
\acmVolume{37}
\acmNumber{4}
\acmArticle{111}
\acmMonth{8}

\begin{document}

\title{Adaptive Human-Agent Teaming: A Review of Empirical Studies from the Process Dynamics Perspective}

\author{Mengyao Wang}
\authornote{Both authors contributed equally to this research.}
\affiliation{%
  \institution{Fudan University}
  \city{Shanghai}
  \country{China}}
\email{mengyaowang23@m.fudan.edu.cn}

\author{Jiayun Wu}
\authornotemark[1]
\affiliation{%
  \institution{Fudan University}
  \city{Shanghai}
  \country{China}}
\email{jiayunwu23@m.fudan.edu.cn}

\author{Shuai Ma}
\affiliation{%
  \institution{Aalto University}
  \city{Helsinki}
  \country{Finland}}
\email{shuai.ma@aalto.fi}

\author{Nuo Li}
\affiliation{%
  \institution{Fudan University}
  \city{Shanghai}
  \country{China}}
\email{linuo@fudan.edu.cn}

\author{Peng Zhang}
\authornotemark[2]
\affiliation{%
  \institution{Fudan University}
  \city{Shanghai}
  \country{China}}
\email{zhangpeng_@fudan.edu.cn}

\author{Ning Gu}
\affiliation{%
  \institution{Fudan University}
  \city{Shanghai}
  \country{China}}
\email{ninggu@fudan.edu.cn}

\author{Tun Lu}
\authornote{Corresponding authors.}
\affiliation{%
  \institution{Fudan University}
  \city{Shanghai}
  \country{China}}
\email{lutun@fudan.edu.cn}

\renewcommand{\shortauthors}{Mengyao Wang et al.}

\begin{abstract}

The rapid advancement of AI, including Large Language Models, has propelled autonomous agents forward, accelerating the \textbf{human-agent teaming (HAT)} paradigm to leverage complementary strengths. However, HAT research remains fragmented, often focusing on isolated team development phases or specific challenges like trust calibration while overlooking the real-world need for adaptability. Addressing these gaps, a process dynamics perspective is adopted to systematically review HAT using the \textbf{T$^4$ framework}: \textbf{T}eam Formation, \textbf{T}ask and Role Development, \textbf{T}eam Development, and \textbf{T}eam Improvement. Each phase is examined in terms of its goals, actions, and evaluation metrics, emphasizing the co-evolution of task and team dynamics. Special focus is given to the second and third phases, highlighting key factors such as team roles, shared mental model, and backup behaviors. This holistic perspective identifies future research directions for advancing long-term adaptive HAT.

\end{abstract}




\maketitle

\section{Introduction}
\label{Section1}

Artificial Intelligence (AI) technologies, exemplified by Large Language Models (LLMs), have enabled the development of autonomous agents that can independently perceive their environment, make decisions, and exhibit goal-directed behaviors across a variety of scenarios. These agents, whether realized as virtual entities \cite{jeon_fashionq_2021} or embodied systems \cite{lin_it_2020}, go beyond traditional algorithms or models by demonstrating higher levels of autonomy and exhibiting diverse degrees of social capability, responsiveness, and proactiveness \cite{luck_conceptual_2001}. In this paper, the term \emph{agent} refers to autonomous agents capable of perceiving, reasoning, and acting within their environment. 



While autonomy is a defining feature of intelligent agents, their purpose is not to operate in complete isolation or fully replace human roles. On one hand, allowing AI to act independently in high-stakes domains raises serious ethical, legal, and accountability concerns—particularly in areas such as healthcare \cite{lee_human-ai_2021} and aviation \cite{zhang_resilience_2023}, where humans must retain ultimate decision-making authority. Even when agents are capable of performing certain tasks autonomously, they must remain under human supervision and defer complex or ambiguous decisions to human experts to ensure responsible and ethical outcomes \cite{madras_predict_2018, lemmer_human-centered_2023}. On the other hand, agents are not perfect in real-world applications. Their performance is constrained by factors such as biased or insufficient training data, difficulty handling out-of-distribution inputs \cite{hemmer_human-ai_2023}, and limited ability to incorporate contextual or tacit human knowledge \cite{muijlwijk_benefits_2024}. These limitations often impair agent performance in complex decision-making scenarios that require situational awareness \cite{laine2023towards}, theory of mind reasoning \cite{strachan2024testing}, or the interpretation of subtle and latent factors. Humans and agents exhibit complementary strengths: agents excel at fast computation and large-scale pattern recognition, while humans contribute contextual understanding, ethical judgment, and adaptive reasoning in uncertain situations.

To capitalize on these complementary strengths and achieve outcomes that surpass the capabilities of either humans or agents alone, the paradigm of \emph{\textbf{human-agent teaming (HAT)}} has emerged \cite{zhang_you_2022, he_interaction_2023, vaccaro2024combinations}. HAT is defined as a collaborative framework in which humans and agents pursue shared goals, distribute responsibilities, and engage in ongoing coordination and negotiation to achieve joint objectives \cite{o2022human, chen2014human, lyons2021human, berretta2023defining}. It has attracted increasing attention from the Human-Computer Interaction (HCI) community as a promising direction for future interactive systems. Researchers have explored HAT by building teams involving one or more humans and agents, and conducting empirical studies with real human feedback to investigate fundamental principles of team formation and collaboration. These studies have addressed aspects such as communication strategies \cite{calisto_assertiveness-based_2023}, collaboration patterns \cite{chiang_enhancing_2024}, and critical issues including trust \cite{kahr_trust_2024, ma_who_2023, ma_are_2024}, bias \cite{rastogi_deciding_2022, zheng_competent_2023}, and various domain-specific applications in healthcare \cite{rajashekar_human-algorithmic_2024}, creativity \cite{singh_where_2023, qin_charactermeet_2024}, programming \cite{ross_programmers_2023, weisz_perfection_2021, ma2025dbox}, and gaming \cite{cimolino_role_2021, flathmann2024empirically}.

However, despite the richness of these studies, we identify two critical gaps in the current HAT literature. \textbf{First}, much of the research remains fragmented, often rooted in legacy HCI topics such as trust calibration. Many studies lack a coherent theoretical foundation and instead propose research questions based primarily on researchers' intuition rather than systematic frameworks. This has led to a piecemeal understanding of HAT, limiting its generalizability and scalability. \textbf{Second}, while adaptability is a key requirement for real-world HAT deployments, it is often underexamined. Teams that can quickly adapt to unforeseen challenges are more likely to thrive and even leverage unexpected opportunities in dynamic environments, whereas those that adapt slowly risk stagnation or failure \cite{terreberry1968evolution}. Yet, many existing studies fail to explicitly address the concept of adaptability or neglect its intrinsic link to the dynamics of team development over time. Therefore, there is a pressing need for a comprehensive and systematic review that not only synthesizes the fragmented findings across HAT research but also provides actionable insights into fostering adaptive team development. Such a perspective is essential for advancing both theoretical understanding and practical implementation of HAT in complex, real-world settings.



To the best of our knowledge, a systematic review of HAT within the Human-Computer Interaction community is still lacking, despite the emergence of related surveys in adjacent fields such as Human-Robot Interaction (HRI). Existing reviews tend to focus on specific facets of HAT, including definitions \cite{lyons2021human, o2022human}, ethical considerations and trust \cite{lopez2023complex}, shared mental models (SMM) \cite{andrews2023role}, team composition \cite{musick2021human}, cohesion \cite{lakhmani2022cohesion}, and self-assessment practices \cite{conlon2024survey}. While valuable, these studies are often narrowly scoped, addressing isolated phases or dimensions of team collaboration without considering HAT as a dynamic and evolving process. As a result, they fall short of offering a comprehensive view of the challenges, complexities, and future directions across the full lifecycle of HAT. Some attempts have been made to introduce integrative frameworks—such as the Input-Mediator-Output (I-M-O) model \cite{o2022human} and the Human-Agent-Team (H-A-T) framework \cite{berretta2023defining}—to organize HAT research more systematically. However, these frameworks typically conceptualize HAT as a predefined, static configuration composed of individual elements (e.g., human, agent, team), with a focus on how inputs are transformed into outputs. This static view overlooks the dynamic, adaptive nature of team development, limiting the applicability of such frameworks to real-world environments where team roles, goals, and interactions continuously evolve. Consequently, there is a critical need for a process-oriented perspective that captures the fluid, adaptive trajectories of HAT over time.

In response to the aforementioned limitations, this paper presents a comprehensive and in-depth review of HAT research within the HCI community through the lens of process dynamics. The goal is to offer researchers a holistic and structured perspective on the full lifecycle of HAT, as well as to provide actionable insights for researchers and designers aiming to build long-lasting and widely adaptive HAT. Drawing inspiration from the theoretical framework for human team development proposed by Kozlowski et al. \cite{kozlowski2008developing}, we introduce the \textbf{HAT Process Dynamics Framework}, referred to as the \textbf{T$^4$ framework}, which comprises four interrelated phases: \textbf{T}eam Formation, \textbf{T}ask and Role Development, \textbf{T}eam Development, and \textbf{T}eam Improvement. This framework conceptualizes HAT not as a static structure, but as a dynamic and evolving process that integrates both task-related and team-development-related trajectories. It provides a unifying lens to systematically organize and synthesize the fragmented body of HAT research within the HCI field. For each phase, we examine the developmental goals, core actions, and corresponding evaluation metrics, identifying the key task-related and social goals that must be achieved to promote stronger team cohesion, adaptability, and long-term effectiveness. As agent capabilities continue to advance—particularly in terms of autonomy, proactiveness, and social intelligence—we envision that agents could increasingly assume leadership or coordination roles within teams. This shift opens the door for HAT to evolve into self-regulating and self-managing entities capable of adaptive behavior in complex, real-world contexts.

Our analysis also reveals that current research efforts are disproportionately concentrated in Phases 2 (Task and Role Development) and 3 (Team Development), focusing on topics such as agent role assignment, coordination and delegation mechanisms, and the development of SMM. In contrast, Phases 1 (Team Formation) and 4 (Team Improvement)—concerning team identity construction and long-term team growth—remain significantly underexplored. This imbalance highlights the need for future research to adopt a team development perspective that spans the full lifecycle of HAT, with greater emphasis on initiating effective teams and sustaining their evolution over time. As an initial step toward bridging this gap, the T$^4$ framework provides a structured, dynamic perspective that connects micro-level interactions with macro-level team development. It serves as a guiding framework for understanding and designing the full cycle of HAT engagement—from initial formation to continuous adaptation—ultimately facilitating more effective, cohesive, and resilient HAT.

The structure of this paper is as follows: Section \ref{Section2} reviews existing survey studies on HAT. Section \ref{Section3} details the methodology for literature search and selection. Section \ref{Section4} introduces the HAT Process Dynamics Framework (T$^4$) and describes the hybrid coding approach used for paper analysis. Section \ref{Section4.1} presents an in-depth examination of two extensively studied phases: Task and Role Development and Team Development. Section \ref{Section5} explores how team development is assessed across different phases. Section \ref{Section6} discusses the application of HAT in both real-world and experimental settings. Section \ref{Section7} outlines future research directions based on the four phases of the T$^4$ framework. Finally, Section \ref{Section8} concludes with a summary of key findings.

\section{Related Work}
\label{Section2}

\renewcommand{\arraystretch}{1.6}
\begin{table*}[htp]  

\fontsize{7}{7}\selectfont  
\caption{Comparison of differences between other reviews and our review. In this table, Analogy denotes "analogy to human teams", and Team Dynamics denotes "Team developmental dynamics", including four phases: team formation, task and role development, team development, and team improvement.}
\label{Tab:comparision-reviews}

\begin{longtable}{cm{5.8cm}cccm{1.5cm}}
\toprule
\textbf{Reviews} & \textbf{Summary} & \textbf{Analogy} & \textbf{Task Dynamics} & \textbf{Team Dynamics} & \textbf{Organizational Structure} \\ \hline

\cite{endsley2023supporting}, 2023 & Examines shared situation awareness in human-AI teams, emphasizing framework development, transparency, and SA-oriented design. & $\times$ & $\checkmark$ & 2, 3 & Framework \\ \hline
\cite{andrews2023role}, 2022 & Reviews SMM in HAT by clarifying definitions, measurement techniques, and relevance, while identifying research gaps and proposing design considerations. & $\times$ & $\checkmark$ & 2, 3 & Framework \\ \hline
\cite{o2022human}, 2022 & Proposes the IMO framework to analyze key variables influencing collaboration in human-autonomy teaming and highlights critical research gaps. & $\checkmark$ & $\times$ & 2, 3, 4 & Framework \\ \hline
\cite{hagemann2023human}, 2023 & Builds on an idealized teamwork process model to explore human-AI collaboration from a team-centered perspective, highlighting key AI capabilities—responsiveness, situational awareness, and adaptive decision-making—and examining technical requirements and challenges. & $\times$ & $\checkmark$ & 2, 3 & Framework \\ \hline
\cite{vats2024survey}, 2024 & Surveys the integration of Large Pre-trained Models with Human-AI teaming by analyzing four dimensions: model improvements, effective HAI systems, safe and trustworthy AI, and applications. & $\times$ & $\times$ & 2, 3 & Framework \\ \hline
\cite{meimandi2024human}, 2024 & Offers a systems-theoretic, interdisciplinary perspective that bridges AI and human-machine interaction, enhancing collaboration, communication, and innovation while addressing socio-technological concerns. & $\times$ & $\times$ & 2, 3 & Framework \\ \hline
\cite{hagos2024ai}, 2024 & Proposes a comprehensive conceptual framework for understanding and advancing AI integration in HAT within tactical environments. & $\times$ & $\checkmark$ & 2, 3, 4 & Framework \\ \hline
\cite{johnson2012autonomy}, 2012 & Examines the impact of autonomy on team performance, particularly when team member interdependencies are inadequately managed. & $\times$ & $\checkmark$ & 2, 3 & Problem \\ \hline
\cite{khakurel2022artificial}, 2022 & Explores how AI can enhance human teams by improving coordination, knowledge sharing, decision support, and overall performance. & $\checkmark$ & $\times$ & 3 & Problem \\ \hline
\cite{berretta2023defining}, 2023 & Investigates the evolution of HAT in response to increasing AI integration in human work, advocating a shift from technology-driven to human-centered approaches that recognize AI as an integral team member. & $\checkmark$ & $\times$ & 3 & Problem \\ \hline
\cite{chen2014human}, 2014 & Examines trust, situational awareness, individual differences, and decision authority in multi-robot control scenarios. & $\times$ & $\times$ & 2, 3, 4 & Factor \\ \hline
\cite{lyons2021human}, 2021 & Highlights social factors and identifies research gaps in Human-Autonomy Teaming. & $\times$ & $\checkmark$ & 3, 4 & Factor \\ \hline
\cite{lakhmani2022cohesion}, 2022 & Explores cohesion differences in teams that include autonomous teammates. & $\checkmark$ & $\times$ & 3, 4 & Factor \\ \hline
\cite{lopez2023complex}, 2023 & Investigates the complex interplay between ethics and trust in human-autonomy teams. & $\times$ & $\times$ & 2, 3 & Factor \\ \hline
\cite{zercher2023ai}, 2023 & Identifies inefficiencies in AI-human team communication and cognition, and proposes directions for future research. & $\times$ & $\times$ & 3 & Factor \\ \hline
\cite{conlon2024survey}, 2024 & Emphasizes self-assessment algorithms that enable autonomous agents to effectively communicate their capabilities. & $\times$ & $\checkmark$ & 3 & Factor \\ \hline
\cite{wischnewski2023measuring}, 2023 & Measures and reviews trust calibrations for automated systems. & $\times$ & $\times$ & 3 & Factor \\ \hline
Ours & Provides a holistic review that conceptualizes HAT through a lens of process dynamics. & $\checkmark$ & $\checkmark$ & 1, 2, 3, 4 & Framework \\

\bottomrule
\end{longtable}
\end{table*}

Understanding the multifaceted characteristics of HAT requires examining its composition, collaboration patterns, and evolutionary mechanisms. Reviews in this field can be classified into three categories based on their focus (Table \ref{Tab:comparision-reviews}): 

\textbf{Problem-Based Reviews.} Focusing on specific challenges,  these reviews are driven by problems in HAT, especially issues affecting team performance. For instance, Johnson et al. \cite{johnson2012autonomy} investigate how autonomy impacts team performance when interdependencies are poorly managed, while Khakurel et al. \cite{khakurel2022artificial} and Berretta et al. \cite{berretta2023defining} explore the evolving role of AI in enhancing team coordination, knowledge sharing, and decision support by shifting from technology-driven to human-centered approaches. \textbf{Factor-Based Reviews.} Concentrating on critical variables, these reviews focus on key factors such as trust, situational awareness, and cohesion. Trust is highly focused on all factors. For instance, Chen et al. \cite{chen2014human} examine trust in multi-robot control, López et al. \cite{lopez2023complex} investigate the interplay between ethics and trust, and Wischnewski et al. \cite{wischnewski2023measuring} explore trust calibrations for automated systems. Besides, Lyons et al. \cite{lyons2021human} highlight social factors, Lakhmani et al. \cite{lakhmani2022cohesion} explore cohesion differences in teams, and Zercher et al. \cite{zercher2023ai} and Conlon et al. \cite{conlon2024survey} address communication inefficiencies and propose self-assessment algorithms for autonomous agents. \textbf{Framework-Based Reviews.} These reviews propose comprehensive models and structured frameworks for HAT, each emphasizing different aspects of team dynamics. For instance, Chen et al. \cite{chen2016human} introduce a three-layer perceptual agent framework focusing on transparency, while Endsley et al. \cite{endsley2023supporting} provide a framework showing the relations of team situational awareness and other factors like transparency. Additionally, Andrews et al. \cite{andrews2023role} present a framework centered on SMM , O’Neill et al. \cite{o2022human} propose the IMO framework, and Hagemann et al. \cite{hagemann2023human} offer a team-centered framework of awareness, information transfer, consolidation, and action.

These reviews provide valuable insights into various factors and issues of HAT and have even structured HAT frameworks. However, they also exhibit limitations. First, they lack a dynamic perspective that organizes HAT's full lifecycle from formation to development to improvement. For example, team roles are not only statically set up after team initialization but also evolve iteratively over time during teamwork, influencing what members of different roles communicate and their self-efficacy. Second, although some reviews propose HAT frameworks by drawing on prior research on human teams \cite{o2022human}, they fall short of incorporating more advanced organizational frameworks and concepts from human team studies. The application of analogical methods remains underexplored. Third, given advancements in technology and trends in HCI, the pursuit of adaptability within the HAT paradigm has become increasingly evident. Meanwhile, research on adaptive human teams has also progressed. However, existing HAT reviews have yet to address this demand or offer guidance on fostering adaptability. 




\section{Methodology}
\label{Section3}

This review adopts a Systematic Literature Review (SLR) methodology \cite{kitchenham2009systematic} to comprehensively synthesize empirical studies on HAT within the Human-Computer Interaction (HCI) community. The primary objective is to establish a solid theoretical foundation and offer practical guidance for advancing the HAT paradigm in real-world applications. Our review process follows established SLR procedures, including: (1) defining the research scope and search terms, (2) selecting relevant databases and publication venues, (3) applying inclusion and exclusion criteria, and (4) conducting multi-stage screening and full-text review.

\subsection{Search Scope and Strategy}

We focused on identifying empirical studies on Human-Agent Teaming (HAT) that involve real human participants, as such studies are essential for understanding the practical dynamics of human-agent teams. To ensure a strong focus on HCI-centered research, we systematically searched leading conferences and journals, including ACM CHI, ACM CSCW, ACM IUI, ACM UIST, ACM TOCHI, PACMHCI, IJHCS, and IJHCI, etc. The core search terms used were “human-agent teaming”, “human-autonomy teaming”, and “human-AI teams”. To broaden our scope and enhance comprehensiveness, we drew inspiration from Kozlowski et al.’s perspective \cite{kozlowski1999developing}, which conceptualizes team growth as a progression from individuals to dyads and ultimately to full team structures. Recognizing their emphasis on dyadic interactions as a fundamental building block for team functioning, we included studies where a single human and a single agent collaborated as a team. Accordingly, we expanded our search terms to incorporate a broader range of terminology: (1) For "agent", we also researched \emph{automat*}, \emph{autonomous}, \emph{AI}, \emph{artificial intelligence}, \emph{LLM}, \emph{large language model}, \emph{model}, \emph{algorithm}, \emph{robot}, \emph{machine}; (2) For "human", we also researched \emph{user}, \emph{participant}, \emph{people}, \emph{individual}; (3) For "teaming", we also searched \emph{collab*}, \emph{team*}, \emph{group}, \emph{cooperation}. This comprehensive search strategy allowed us to include studies that may not explicitly use the term “HAT” but investigate collaboration mechanisms between humans and intelligent agents—such as decision-support systems, cooperative workflows, and interactive AI systems—thus contributing meaningfully to the development of HAT.

\subsection{Screening and Selection Process}

\begin{figure}
    \centering
    \includegraphics[height=4.5cm, width=15.0cm]{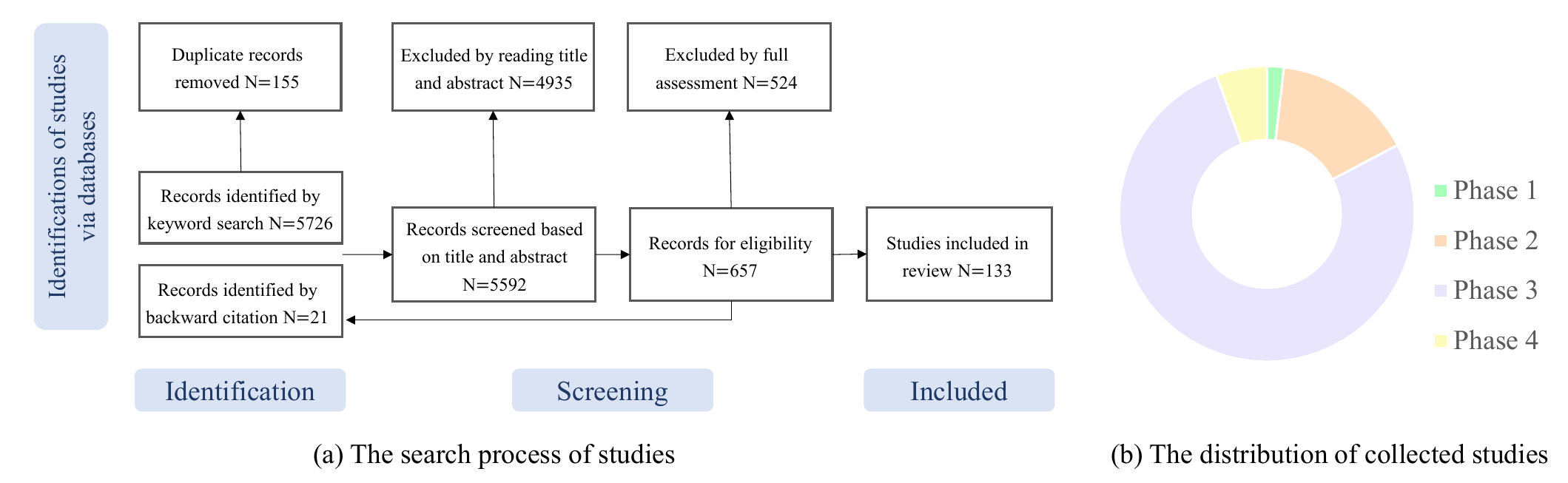}
    \caption{(a) Overview of the paper search and inclusion process following the Systematic Literature Review (SLR) methodology. (b) Distribution of selected papers, with a concentration in Phase 2 and Phase 3.}
    \label{fig:search}
\end{figure}

The search initially yielded approximately 6,000 papers. We first removed duplicates and then conducted a title and abstract screening to exclude irrelevant studies. Next, we performed a full-text review to assess whether each study met our inclusion criteria: (1) The presence of real human participants, ensuring that the study provides insights into human-agent interaction rather than focusing solely on technical model advancements. (2) A focus on human-agent collaboration or teaming, specifically involving humans working directly with relatively autonomous agents toward a shared goal, instead of just humans using an AI-supported tool. (3) Use of empirical methodology (e.g., lab studies, field deployments, interviews), allowing for an examination of humans' actual perceptions in HAT. (4) Publication in an HCI-related venue. The inclusion process was conducted collaboratively by two authors. Any disagreements were resolved through discussion, and a weekly meeting with all authors was held to review and refine the inclusion decisions. After applying these criteria, a total of 133 papers published between 2007 and 2024 were selected for final inclusion. This corpus captures both the historical development and recent advancements in HAT research within the HCI community. To contextualize our analysis, we also examined relevant HAT-related papers from adjacent technical venues such as NeurIPS, AAAI. However, as many of these works lack solid empirical engagement with human participants, they were not included in searching. Instead, we incorporated them through one-round backward citation analysis, maintaining consistency with the SLR search protocol (see Fig. \ref{fig:search}).



\section{The T$^4$ Framework of Modeling HAT Process Dynamics and Coding Method}
\label{Section4}
In this section, we first introduce the HAT Process Dynamics Framework, T$^4$ (\textbf{T}eam Formation, \textbf{T}ask and Role Development, \textbf{T}eam Development, and \textbf{T}eam Improvement). We then apply this framework to code and categorize the collected papers through a hybrid approach of inductive coding and deductive coding.


\subsection{The T$^4$ Framework of HAT Process Dynamics}

To understand how HAT evolves and adapts in dynamic environments, we draw inspiration from Kozlowski et al.'s \cite{kozlowski2008developing} theory of adaptive human team development, which describes how teams refine roles, transition across phases, and achieve dynamic leadership in response to changing contexts. This perspective aligns with HATs, where team identity, role distribution, and coordination emerge fluidly rather than following a fixed sequence, adapting to real-world complexities. For instance, Sidji et al. \cite{sidjiHiddenRulesHanabi2023} highlight the need of the dynamic role negotiation among humans and agents in the Hanabi game. Moreover, the necessity for HATs to operate in uncertain, evolving environments further reinforces their alignment with Kozlowski et al.'s theory, emphasizing both the objective of adaptation and the dynamic nature of the process. A particularly strong parallel between human teams and HATs lies in the evolution of leadership. Kozlowski et al. suggest that team improvement can lead to the development of a compatible mental model, where leadership is not rigidly assigned but instead distributed dynamically. As agents gain autonomy, a similarly flexible mechanism to leadership becomes increasingly necessary, reducing human workload and enabling more efficient collaboration. 

\begin{figure}
    \centering
    \includegraphics[height=6.64cm, width=15.5cm]{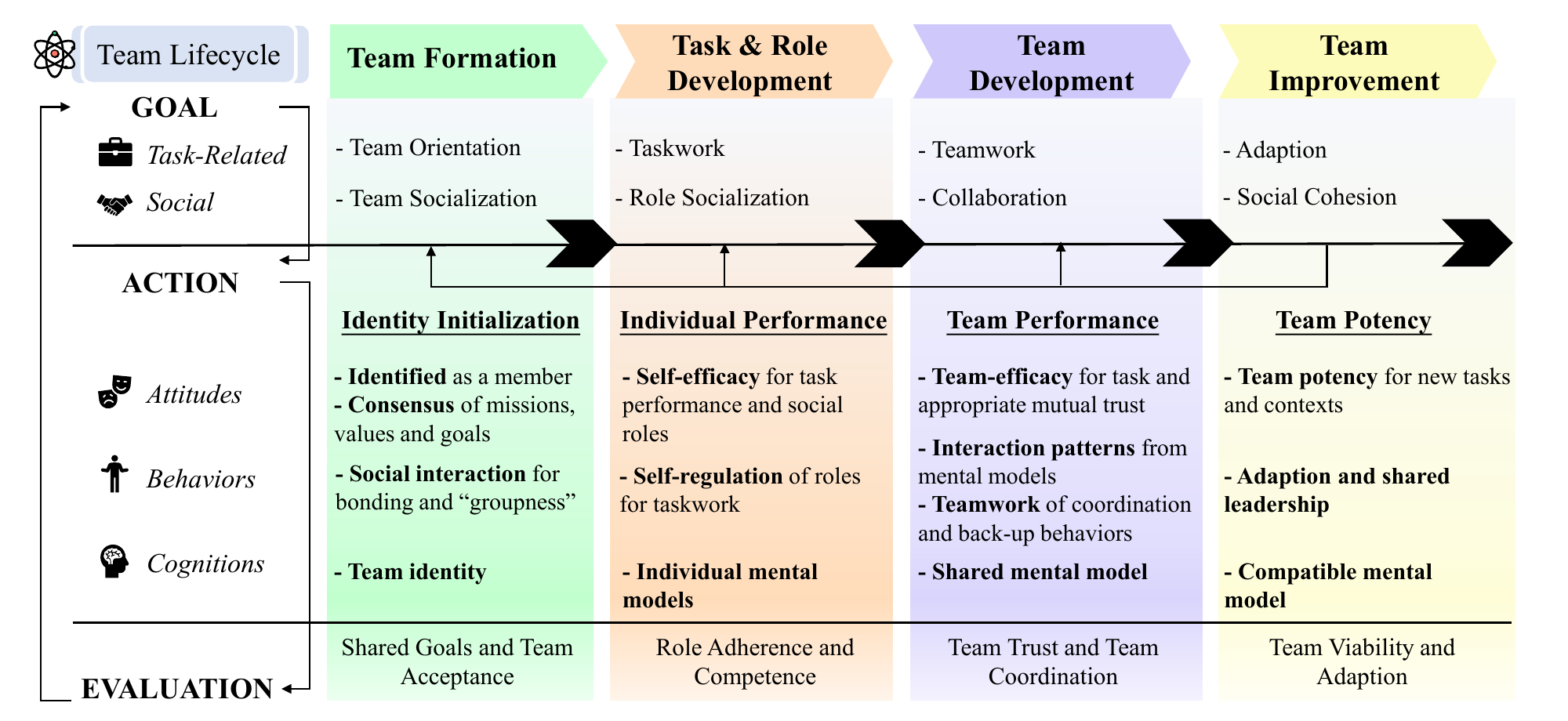}
    \caption{T$^4$ framework of HAT process dynamics. The left side illustrates the goal-action-evaluation cycle of task dynamics, which takes on different meanings at each phase of team developmental dynamics: \colorbox{green!10}{Team Formation} focuses on establishing team identity; \colorbox{orange!10}{Task and Role Development} emphasizes role assignment and the task execution; \colorbox{blue!10}{Team Development} is the critical phase, centering on teamwork and collaboration; \colorbox{yellow!10}{Team Improvement} addresses long-term sustainability and adaptability. The interplay between task dynamics and team developmental dynamics forms a process dynamics perspective spanning the entire HAT lifecycle.}
    \label{fig:framework}
\end{figure}

However, unlike human teams, where leaders actively shape development through strategic interventions, leadership in HATs is constrained by the technical nature of AI training and deployment, limiting direct human oversight. To better suit the HAT context, we restructure task dynamics into a goal-action-evaluation cycle applicable to both humans and agents. Additionally, we refine certain aspects of the original framework to align more closely with the conventions of the HCI and AI communities. Ultimately, we introduce the \textit{T$^4$ Framework of HAT Process Dynamics} (Fig. \ref{fig:framework}). This framework views HAT development as a dynamic process, driven by two key dynamics:
\begin{itemize}
    \item \textbf{Task dynamics}, which describe the cyclical process through which team members set goals, execute tasks, evaluate outcomes, adjust strategies, and acquire new skills in each iteration. 
    \item \textbf{Team developmental dynamics}, which describe the iterative and recursive progression of a team, consisting of team formation, task and role development, team development, and team improvement.
\end{itemize}

These two dynamics are interwoven. On one hand, through the cycle of task dynamics, members achieve targeted goals in the current phase of team development, advancing to the next phase \cite{kozlowski2008developing}. On the other hand, each development phase imposes distinct requirements on the cycle of task dynamics elements such as goals and actions. Ultimately, the team strives toward becoming a more self-managing and self-regulating entity capable of adaptation.


In the \colorbox{green!10}{team formation} phase, the team's development goal is to initialize team identity. Members must identify with the team and be willing to contribute to it, reaching a shared understanding of the team’s mission, norms, and values \cite{ellemers1998group}. Members need to engage in social interactions to know each other, build connections, and form a sense of “groupness.” Therefore, the social ability of team socialization is a key development goal. Since no specific tasks have truly begun at this phase, the task-related goal is simply to ensure that members have some understanding of “what do I need to do in the team” and “what can I do to help accomplish the mission,” so new members are oriented to reduce ambiguity \cite{b_tuckman_developmental_1965}. The development evaluation for this phase, therefore, focuses on building team identity, namely shared goals of the team and team acceptance.

In the \colorbox{orange!10}{task and role development} phase, the team's development goal is to improve members' individual task capability and their respective roles on the team \cite{ostroff_organizational_1992}, so the task-related goal is quite straightforward, while the social goal emphasizes building social role acceptance and attachment. Members need to develop self-efficacy, a belief in their abilities \cite{butler1998self}, and self-regulation, a metacognition that involves planning, monitoring, and modifying their cognition and behaviors. Therefore, the development evaluation for this phase focuses on role adherence (whether roles are followed), and competence (whether roles are performed competently).

In the \colorbox{blue!10}{team development} phase, the team's development goal is to improve teamwork, meaning members should know when, which member, and what to provide \cite{kozlowski1999developing}. This is also a key aspect of the task-related goal. Additionally, from the perspective of social development goals, members need to genuinely cooperate to stimulate positive interactions. Therefore, members have appropriate trust, responsible reliance, and mutual respect, which together build team efficacy \cite{gully_meta-analysis_2002}. Under such teamwork, team members can construct a certain level of SMM and use it to guide interaction patterns, coordinating with each other and providing back-up behaviors, which is also the focus of team evaluation.

Finally, in the \colorbox{yellow!10}{team improvement} phase, the development goal is to become an adaptive HAT capable of thriving in diverse and dynamic environments. From a task-related perspective, the team should be able to adapt to more diverse, complex, and novel task scenarios \cite{kozlowski1999developing}. From the social perspective, the team should foster social cohesion \cite{barrick1998relating} , where members identify each other more, ensuring the team's long-term persistence. Additionally, the team should address critical issues such as workload balancing and conflict management, extending beyond phase 3. With an emphasis on the team's potency for new tasks and contexts, members further consider the possibility of shared leadership, leading to the construction of compatible mental models \cite{pearce_role_2003}. However, this extends beyond much of the existing HAT research, and even human teams rarely reach this phase. Thus, it is considered a key research direction with implications for the design of HATs. Accordingly, the evaluation of this phase focuses on team viability (whether the team can sustain itself) and adaptability (whether it can effectively adjust to changing conditions).



\subsection{Coding Papers through a Hybrid Approach}

We used a hybrid inductive and deductive coding approach \cite{feredayDemonstratingRigorUsing2006} to analyze 133 papers. First, we developed an initial code manual based on the T$^4$ framework, mapping key task dynamics to specific phases. For example, “shared mental model” was identified in phase 3 as a key cognitive component. To ensure consistency, two authors independently coded a subset of papers in a pilot phase, resolving discrepancies through discussion and refinement. For instance, the term “commitment” varied across phases, so we clarified it in phase 1 as “acceptance of mission, values, and goals” \cite{ellemers1998group}, and in phase 3 as “commitment”, part of shared mental models \cite{hodhod_closing_2016}. Following refinement, we summarized coding results and identified themes like the roles of “advisor” and “challenger” in phase 2. We then conducted large-scale coding, introducing additional codes as needed, such as consolidating “error”, “accuracy”, “recall”, and “precision” into “effectiveness”. New codes were validated by all coders before applying them to all papers. After coding, we synthesized codes into higher-level themes, like “implementer” and “coordinator” in phase 2, and “perception gaps” in phase 3. Finally, all authors reviewed and validated the themes and coding consistency.

\section{T$^4$ Lifecycle Framework}
\label{Section4.1}
The coding results show that phases 2 and 3 are the focal points of HAT research (Fig. \ref{fig:search}). Therefore, in this section, we focus on reviewing research related to these two phases, analyzing key research points concerning HAT, while in Section \ref{Section6}, we discuss research on phases 1 and 4, identifying future directions. Additionally, we believe that a comprehensive review of this phase across all four development phases is beneficial for identifying gaps in HAT team development and serving as a reflection for improving HAT design. To this end, we have collected relevant evaluation metrics from empirical experiments and discuss them from the perspective of the four development phases in Section \ref{Section5}. 

\subsection{Task and Role Development}
\label{Section4.2}
The key aspect of this phase is defining HAT roles, which focus on individual members—what roles they assume, what tasks they need to accomplish, and how they develop self-efficacy regarding their roles and tasks. Given that HATs differ from human teams and agents require additional design, we specifically focus on the roles that agents play within HATs. Additionally, we provide a reference for the structure of HATs.

\subsubsection{Structure of HAT}
In terms of team composition, humans and agents serve as core members, each with unique design and role characteristics: \textbf{For humans}, a fundamental characteristic is demographic information, such as age \cite{ho_its_2024}. Moreover, expertise is often considered, particularly in specialized HATs involving human experts \cite{callaghan_mechanicalheart_2018, muijlwijk_benefits_2024, wang_lave_2024} or novices \cite{callaghan_mechanicalheart_2018, kariyawasam_appropriate_2024, overney_sensemate_2024}. \textbf{For agents}, their design and configuration are inherently more complex and technical. Thus, their characteristics expand across embodiment \cite{shamekhi_face_2018} and more subtle characteristics like explainability \cite{feng_what_2019}, perceived identity \cite{hwang_ideabot_2021}, tone of communication \cite{calisto_assertiveness-based_2023} and so on.

The composition of HATs primarily focuses on the ratio of agents to humans, forming four structural types: \textbf{1v1}, \textbf{1vN}, \textbf{Nv1}, and \textbf{NvN}. In the \textbf{1v1} structure, a single agent interacts with a single human. This structure is commonly found in collaborative systems where the agent has relatively low autonomy, such as the qualitative coding system \cite{gebreegziabher_patat_2023}. Additionally, many studies adopt this structure as the minimal configuration of HATs to explore the impact of various characteristics, such as ambiguity-aware explanation \cite{schaekermann_ambiguity-aware_2020}. In the \textbf{1vN} structure, a single agent interacts with multiple humans. This configuration is frequently examined for exploring how agents influence human teams, such as empowering certain members \cite{taylor_coordinating_2019} or mediating conflicts \cite{zheng_competent_2023}. In the \textbf{Nv1} structure, multiple agents interact with a single human. This configuration is relatively uncommon, and existing research mainly treats multiple agents as part of the technological infrastructure \cite{trajkova_exploring_2024} rather than independent team members. In the \textbf{NvN} structure, multiple agents interact with multiple humans. This configuration is typically found in purposefully designed HATs, though research on this structure remains limited. For instance, Jung et al. \cite{jung_engaging_2013} designed a team comprising two humans and three agents to examine the impact of backchanneling on teamwork.

\begin{figure}
\centering
\includegraphics[width=0.5\linewidth]{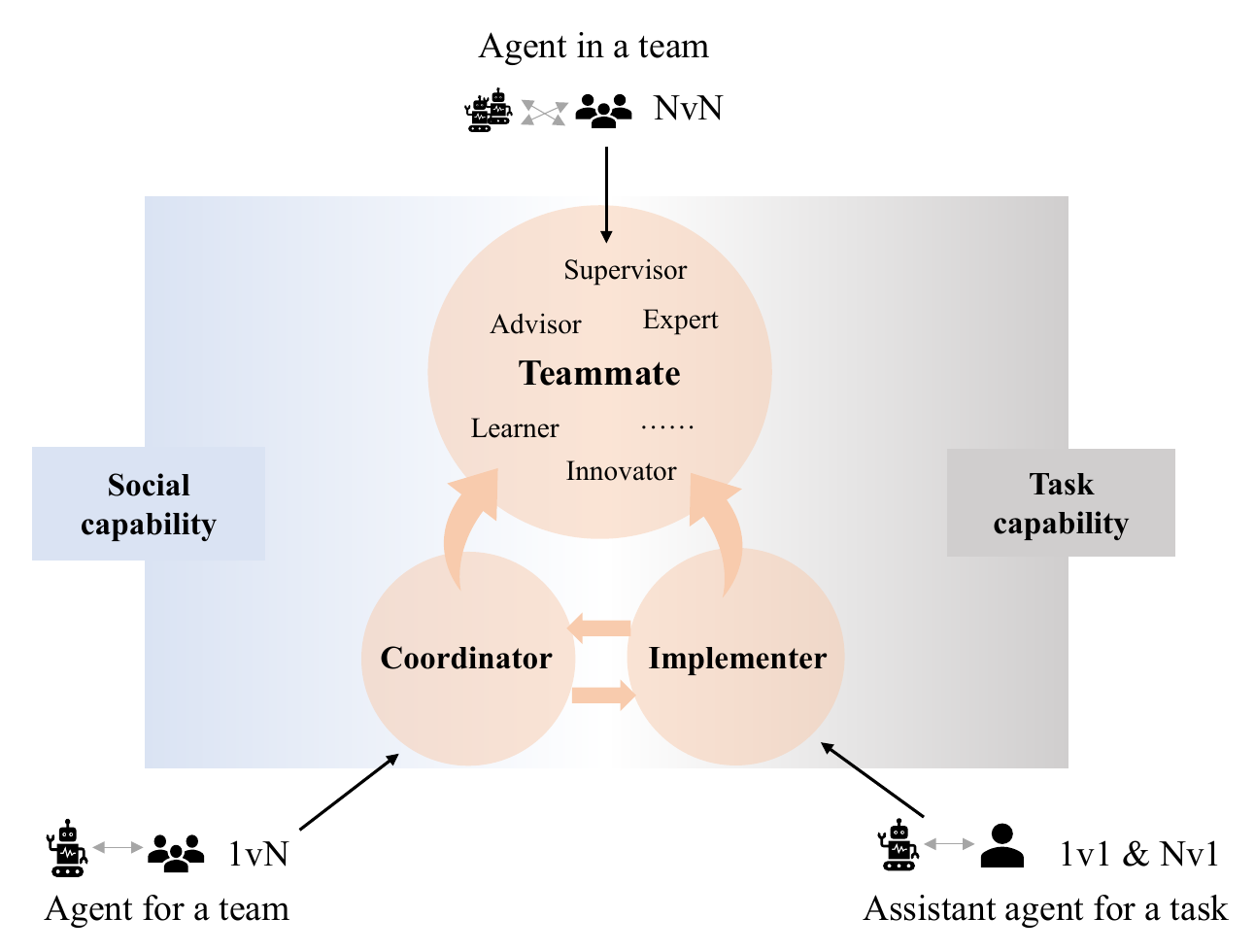}
\caption{Capabilities and roles of agents in HAT. Existing research on agent roles in HAT follows two main threads: (a) Agent for a Team, typically in a 1vN structure, where agents act as coordinators with a focus on social capability; (b) Assistant Agent for a Task, often in 1v1 or Nv1 structures, where agents serve as implementers, emphasizing task capability. As agents gain greater autonomy, they should evolve beyond these categories to dynamically balance social and task capabilities, engaging as versatile teammates in HAT with diverse roles.}
\label{fig:agent roles in HAT}
\end{figure}
\subsubsection{Roles of Agents in HAT}
The roles of the agents in HAT define their responsibilities and are crucial to the formation of organizational architecture and individual performance \cite{wu2019agent,zheng2019roles,shi_agent_2013}. Drawing on the experience of human teams, where technical and social skills drive success \cite{stevens1994knowledge,robbins2018essentials}, we propose two fundamental capabilities of agents: \textbf{Task Capability,} the ability to execute specific tasks, including requirement understanding, planning, execution, evaluation, and strategy refinement. A strong task capability directly enhances efficiency, effectiveness, and task success rates. \textbf{Social Capability,} the ability to interact effectively with humans and other agents through information exchange, conflict resolution, and team coordination. These capabilities align with the task-related and social goals\textbf{ }across different phases of HAT development. Based on these dimensions, we identify two fundamental roles: \textbf{Implementer and Coordinator}—each emerging from distinct team structures and evolving to support human-agent collaboration.

\textbf{Implementer: Task-Oriented Agents in 1v1 and Nv1 Structures}. The implementer role originates from 1v1 and Nv1 structures, where agents primarily serve as assistants for individual tasks. In these configurations, agents focus on task execution, leveraging their task capability to efficiently complete assigned responsibilities. For example, in autonomous driving \cite{trabelsi_advice_2023} and medical diagnostic assistance \cite{cai_human-centered_2019,rajashekar_human-algorithmic_2024}, agents act as task-oriented partners, providing real-time support and decision-making assistance to individual users. With advancements in AI technology \cite{kuang_enhancing_2024,taigman2014deepface}, implementations not only execute instructions efficiently but also understand task contexts and proactively propose better solutions \cite{zhang_resilience_2023}. These agents translate human ideas into tangible results, offering instant responses and assistance, as seen in user experience evaluation \cite{kuang_enhancing_2024}, creative design \cite{dhillon2024shaping,yuan_wordcraft_2022,wang_lave_2024,ross_programmers_2023}, and clinical environments \cite{wang_brilliant_2021}.

\textbf{Coordinator: Team-Oriented Agents in 1vN Structures}. The coordinator role evolves from 1vN structures, where agents collaborate with multiple humans to support team communication and coordination. In these configurations, agents emphasize social capability, facilitating information exchange, resolving conflicts, and optimizing workflows. For example, in co-creation tasks such as programming, writing, and crowdsourcing, agents act as team coordinators, enhancing collaboration among human members \cite{he_towards_2024}. By acting as mediators or "social glue" \cite{suh2021ai}, coordinators improve team dynamics and psychological safety, enabling teams to operate more efficiently. They support decision-making and problem-solving, ranging from parallel editing \cite{hymes_unblocking_1992} to complex qualitative analysis \cite{gao_collabcoder_2024}.

\textbf{Specialized Roles: Integration of Implementers and Coordinators}. While HAT roles mainly fall into Implementers and Coordinators, specific tasks often require refined or hybrid roles. For example, military simulations adapt roles to dynamic missions, while industrial automation uses agents for real-time data sharing and task allocation, reflecting nuanced applications of these core roles \cite{zhang_you_2022}. Four specialized roles—\textit{Advisors}, \textit{Supervisors}, \textit{Innovators}, and \textit{Learners}—extend the core functions of Implementers and Coordinators, often blending both. \textit{Advisors} provide expert guidance, typical of Coordinators, to navigate complex scenarios, such as enhancing diagnostic accuracy in medical teams \cite{rajashekar_human-algorithmic_2024}. \textit{Supervisors} oversee workflows, balancing implementation and coordination, as seen in creative design projects ensuring consistency and quality \cite{chiang_enhancing_2024}. \textit{Innovators} drive new ideas and strategies, aligning with the Implementer role \cite{karimi_creative_2020}. \textit{Learners} evolve through interaction and feedback, adapting to tasks while integrating implementation and coordination \cite{feng_what_2019}. These four specialized roles are not distinct from the core Implementer and Coordinator categories but rather context-specific extensions that often integrate both. Future research should examine how these roles can be systematically incorporated into team dynamics to strengthen human-agent collaboration.


\subsection{Team Development}
\label{Section4.3}


At this phase, HAT truly develops into a team, aiming for improved teamwork in task-related goals and enhanced collaboration in social aspects, such as fostering appropriate trust and reliance among members. A key aspect of this development is team efficacy—a shared belief among team members regarding the team’s capability to perform the task effectively. These objectives are closely linked to the SMM in cognition. Therefore, we first explore how the literature addresses the construction of SMM for HAT, examining the nature of their perpetual negotiation and the challenges they face. In particular, we identify factors like inappropriate beliefs as obstacles in this process, highlighting a broader research design space. While SMMs shape cognitive alignment, team members exhibit diverse interaction patterns at the behavioral level to enhance teamwork. Therefore, we review how coordination and mutual support are achieved in HAT to inform the design of specific interaction patterns within HAT.

\subsubsection{'Perpetually Negotiating': Building an SMM of HAT}

\begin{figure}
    \centering
    \includegraphics[width=0.95\linewidth]{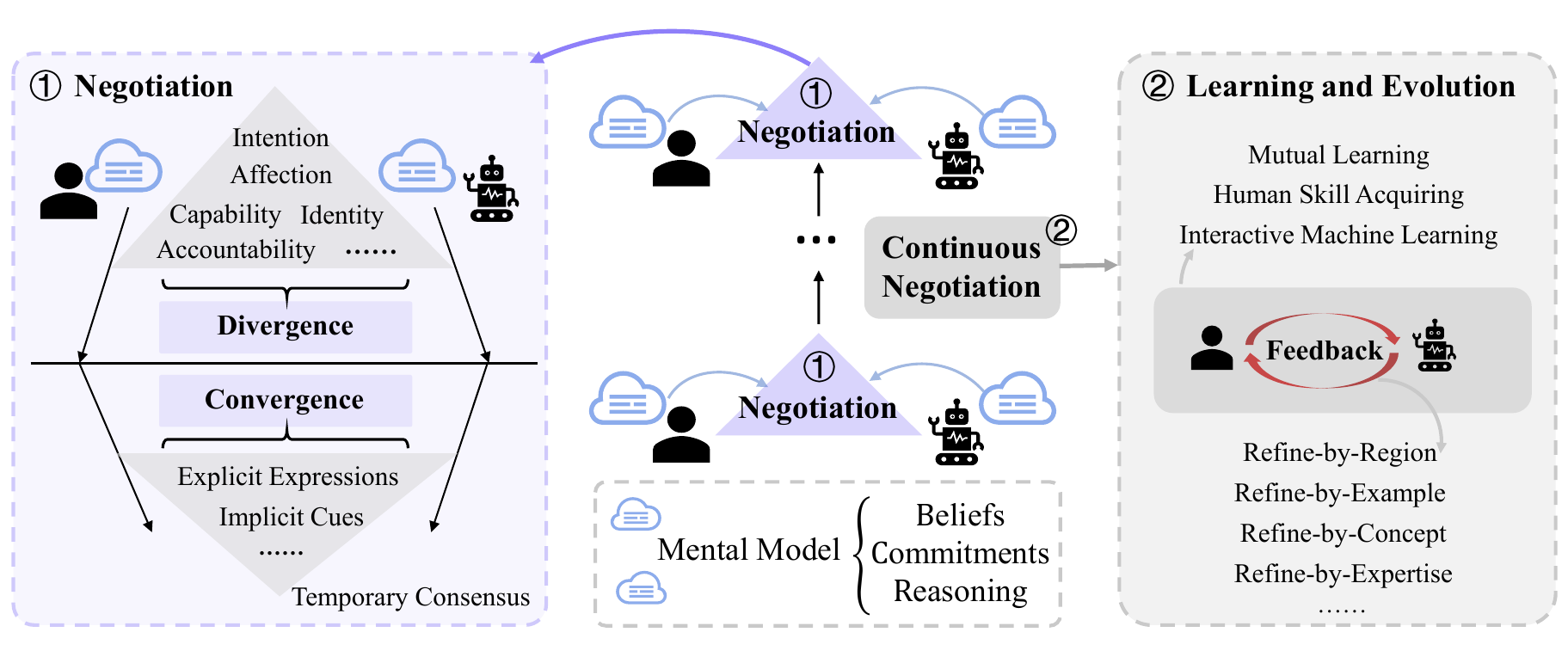}
    \caption{Perpetual Negotiation of SMM. The \textbf{left} side illustrates the negotiation processes in constructing an SMM, including divergence and convergence, where divergence involves information exchange, and convergence highlights how explicit expressions and implicit cues—particularly the latter—aid in reaching temporary consensus. The \textbf{right} side demonstrates how continuous feedback fosters mutual learning between humans and agents, emphasizing the design mechanisms that enable agents to evolve throughout the continuous negotiation process.}
    \label{fig:negotiate}
\end{figure}

The concept of shared mental model is closely linked to Theory of Mind (ToM), which enables individuals to interpret and predict others' behaviors based on both observable and latent cues \cite{wang_towards_2021}. In human teams, members continuously adapt to each other through interaction, influencing thoughts and behaviors over time \cite{flathmann2024empirically}. When autonomous agents join the team, a similar process occurs: as agents interpret users' inputs and adjust accordingly, they also shape users' thinking in return, demonstrating mutual adaptation \cite{lin_it_2020}. This reciprocal understanding between humans and agents forms a \textbf{Mutual Theory of Mind (MToM)}, laying the foundation for \textbf{SMM} in HAT. SMM refers to the collectively developed shared mental representations that enable team members to coordinate effectively \cite{park_retrospector_2023, wang_towards_2021}. However, unlike static models, SMM is not pre-defined but perpetually negotiated. Similar to improvisational performances, where individuals engage in "perpetual negotiation"\cite{hodhod_closing_2016}, humans and agents in HAT engage in an ongoing exchange of cues, leading to divergence and convergence of mental models over time (Fig. \ref{fig:negotiate}).

\textbf{Negotiation: Divergence and Convergence.} Divergence refers to the content of negotiation, ranging from global to local information. Global aspects like identity\cite{hwang_ideabot_2021} and capability\cite{wang_lave_2024} change rarely or only in long-term interactions, while relatively local factors such as affection and engagement\cite{kim_engaged_2024} shift in short-term interactions. Local information, like intention, may vary with each interaction. Negotiation methods then converge this information to a temporary consensus that guides behavior. These methods are either explicit, involving clear communication like predictions (not the focus here), or implicit, defined as channels ranging from non-explicit verbal statements to other means as subtle as eye gaze or gesture\cite{liang_implicit_2019}, which encompass implicit communication\cite{zhang_resilience_2023} and observable behavioral cues\cite{jung_engaging_2013, shamekhi_face_2018}. Subtle verbal cues align with the concept of implicature\cite{grice1978further}, a cooperative speech. Liang et al.\cite{liang_implicit_2019} apply this in Hanabi, finding that agents using implicature were perceived as more human-like. Similarly, Zhang et al.\cite{zhang_resilience_2023} explore implicit suggestions in decision support systems for aviation, where hints subtly facilitate understanding. Observability\cite{christoffersen2002make} of behavioral cues enhances teaming by conveying either task-related\cite{mccormack_silent_2019} or collaboration-related information\cite{jung_engaging_2013, shamekhi_face_2018}, the latter of which is referred to as backchanneling, and help coordinate interactions and signal understanding, increasing social presence, perceived capability, etc. However, agents, despite mimicking human behaviors, may evoke different perceptions due to their unique identities. Kim et al.\cite{kim_engaged_2024} find that while agents can foster engagement, they may also make interactions feel more competitive. This highlights that while implicit negotiation through backchanneling can enhance teaming, it requires careful consideration to avoid unintended outcomes.

\textbf{Learning and Evolution.} Through long-term negotiation of mental models, humans and agents develop reciprocal representations that align expectations and enhance performance via mutual learning \cite{ross_programmers_2023}. Although short-term collaboration can lead to local adaptation, such as adapting to specific data or tasks \cite{gebreegziabher_patat_2023}, we focus on global learning and evolution that emerge over time as part of the construction of the SMM. Human learning in HATs often revolves around acquiring specific skills, such as programming \cite{kuttal_trade-offs_2021} or reading \cite{ho_its_2024}. In these cases, agents function more as tutors than teammates, guiding humans in skill development and leading to an asymmetric HAT structure. Additionally, existing research predominantly focuses on local learning or evaluation metrics like self-efficacy, neglecting broader human learning dynamics\cite{gebreegziabher_patat_2023, hwang_ideabot_2021}. In contrast, agent learning in HATs requires more sophisticated design considerations. Interactive Machine Learning (IML) frames model training as a human-computer interaction (HCI) task \cite{overney_sensemate_2024}, allowing humans to actively shape agent learning through interaction and feedback \cite{muijlwijk_benefits_2024}. IML applications range from engaging novices in model development \cite{mohammadzadeh_studying_2024} to leveraging expert knowledge in critical domains like healthcare \cite{cai_human-centered_2019, lee_human-ai_2021}. A key focus is the design of the training loop, where feedback mechanisms such as refine-by-region, refine-by-example, and refine-by-concept help agents align their representations with human mental models \cite{cai_human-centered_2019}. Furthermore, iterative refinement in IML supports personalization, enhancing agent adaptation over time \cite{jahanbakhsh_exploring_2023, ruoff_onyx_2023, shi_retrolens_2023, lee_human-ai_2021}.

\textbf{Agents as SMM Coordinators.} In certain HATs, agents act as coordinators (see Section \ref{Section4.2}), indirectly contributing to task completion by supporting the development of SMM\cite{zheng_competent_2023}. They address challenges in human teams, such as evaluation apprehension (hesitation due to fear of judgment) and free riding (unequal participation), which stem from visible contributions and power dynamics \cite{hymes_unblocking_1992}.  Typically engaging in a 1vN structure of HAT, agents help mitigate these issues by fostering playfulness and reducing social pressure, acting as a 'common foil' to encourage creative risk-taking\cite{suh_ai_2021}. This fosters a judgment-free environment, improving idea generation\cite{hwang_ideabot_2021}. Additionally, agents contribute through social efforts such as establishing common ground, breaking the ice, mediating conflicts, and empowering less dominant team members\cite{shamekhi_face_2018, kim_moderator_2021, taylor_coordinating_2019, gao_collabcoder_2024}. Though agents cannot eliminate power dynamics, they can promote collaborative outcomes\cite{gao_collabcoder_2024}. In addition, agents like social robots also facilitate interactions through their embodied presence, particularly in triadic relationships, like those between healthcare robots, staff, and older adults, forming an SMM among all parties\cite{ho_its_2024, yuan_social_2022}.

\subsubsection{Gap Between Real and Perceived Worlds: Challenges in Constructing an SMM}


The design space for constructing SMMs encompasses multiple sub-design spaces. For example, Zheng et al. \cite{ehsan2021expanding} propose a process-oriented view within XAI, arguing that XAI should convey both model-related information and task-related information. This perspective implicitly suggests a more interactive design space—one where users and AI systems actively align through task-based information exchange. Moreover, Yang et al. \cite{yang_harnessing_2023} highlight that simply “exploring how AI works” is essential for issues like AI accountability but does not directly influence user trust. As a result, they advocate for shifting the focus from "XAI" to trust-calibration interaction design, introducing a broader design space that explores diverse mechanisms and interactions for calibrating trust. We argue that \textbf{the SMM design space is even broader}, integrating and extending beyond these existing design spaces (such as XAI and trust calibration) to address deeper cognitive challenges. To grasp this space, we must revisit the definition of mental models—mental representations used to predict a subject's behavior in the world \cite{he_interaction_2023}. Therefore, constructing an SMM involves creating accurate representations and effective predictions, which reveals two main gaps: the gap between how team members perceive the world and the real world (perception gap), and the gap between how team members perceive the world and how they can express this understanding (expression gap). Building an SMM is fundamentally about bridging these gaps, with many studies implicitly addressing the first gap—the \textbf{perception gap}. Given the additional design considerations for agents' mental models and their inherent black-box nature, much of the research has focused on calibrating human perceptions of the real world. This remains the primary focus here.


\begin{figure}
    \centering
    \includegraphics[width=0.78\linewidth]{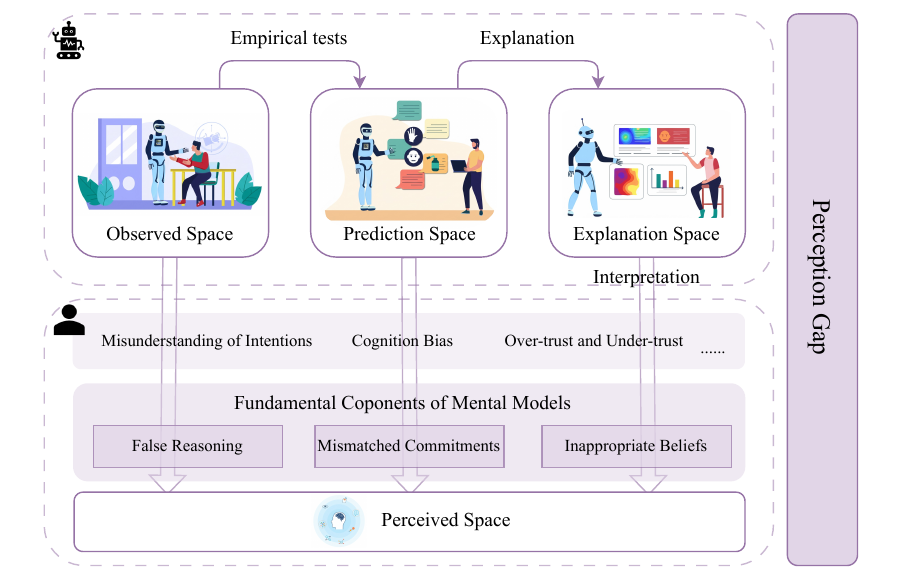}
    \caption{The perception gap between real world and perceived world. This figure illustrates the perception gap between the perceived space and three other spaces: observed space, prediction space, and explanation space. This gap arises from fundamental components of mental models—false reasoning, mismatched commitments, and inappropriate beliefs—which lead to cognitive biases, misplaced trust, and misunderstandings etc. This perception gap poses a significant challenge in constructing an SMM, while its calibration subsequently influences interaction patterns like reliance.}
    \label{fig:Gap}
\end{figure}

To illustrate this perception gap, we extend Rastogi et al.'s model of cognitive biases\cite{rastogi_deciding_2022}. Agents operate within an observed space to make predictions, forming a prediction space. These predictions, along with explanations such as confidence levels, constitute the explanation space. Humans, in turn, observe the world, receive predictions and explanations, and construct a perceived space. However, a gap remains between the human-perceived world and the real world (namely the three other spaces). This gap arises from three fundamental components of mental models: \textbf{false reasoning, mismatched commitments, and inappropriate beliefs}  \cite{hodhod_closing_2016}. Prior efforts, such as trust calibration and cognitive therapy, are not isolated solutions but rather address interwoven aspects of these components, forming a key part of the SMM design space. By clarifying this gap, we provide insights into constructing SMMs and fostering the cognitive development required for HAT to progress to Phase 3. We further explore three key challenges within this gap and how calibration of the gap influences interaction patterns (shown in Fig. \ref{fig:Gap}).

\textbf{False Reasoning.} The reasoning process influences perception, shaped by the quality of information and the reasoning path. Explanations should provide useful, timely information, such as natural language rationales for non-experts\cite{das_leveraging_2020}, literature-backed evidence for experts\cite{yang_harnessing_2023}, or clarifying arguments for ambiguous cases\cite{schaekermann_ambiguity-aware_2020}. Specialized mechanisms like AI-framed questioning\cite{danry_dont_2023} and counterfactual explanations\cite{lee2023understanding} enhance reasoning. Additionally, explanation factors such as assertiveness\cite{morrison2024impact, calisto_assertiveness-based_2023} and correctness\cite{morrison2024impact} affect how users perceive the information. Reasoning is shaped by both objective factors like cognitive load\cite{he_knowing_2023} and reasoning ability\cite{pinski_ai_2023}, and subjective factors like emotions and biases. For instance, algorithm aversion\cite{he_knowing_2023} can lead individuals to reject correct reasoning. Zheng et al.\cite{zheng_competent_2023} note that participants' biases against AI, such as stereotypes, may stem from unfamiliarity, putting the reasoning process at a disadvantage from the start.


\textbf{Mismatched Commitments.} Collaboration perceptions are shaped by awareness of personal and others' obligations. People often conflate commitments with beliefs, as their expectations of an agent's competence inform its perceived responsibilities. Conversely, expectations influence perceived competence, which is our focus. Assigning distinct roles helps manage responsibilities—Ashktorab et al.\cite{ashktorab_effects_2021} find that agents in a 'giver' role were seen as more intelligent in a word-guessing game. This may be due to attention focused on the agent's specific task performance, leading to an overly optimistic evaluation of its overall intelligence, referred as a violation of the choice-independence assumption\cite{pinski_ai_2023}. Agents' behaviors also signal implied commitments. Clark et al. \cite{clark_creative_2018} noted that more intrusive systems encouraged interaction but raised expectations for valuable advice, risking disappointment. To mitigate this, Arakawa et al. \cite{arakawa_catalyst_2023} avoided active engagement buttons to prevent unrealistic expectations. This tradeoff highlights that while anthropomorphic designs enhance engagement, over-reliance on human social norms can lead to unmet expectations \cite{sadeghian_artificial_2022}. Given current agent limitations, a restrained design with clear role definitions is crucial. Blurred responsibility divisions can also lead to human free-riding \cite{hymes_unblocking_1992, shi_retrolens_2023, gao_collabcoder_2024}. As humans still lead teams, managing reliance remains a challenge. Future agents should help clarify human commitments and encourage active participation, emphasizing human initiative.

\textbf{Inappropriate Beliefs.} Beliefs often relate to trust calibration, defined as aligning people's trust in AI with its actual capabilities\cite{ma_who_2023}. This calibration can focus both self-beliefs and beliefs about agents. He et al.\cite{he_knowing_2023} discuss the Dunning-Kruger Effect, where less competent individuals overestimate their abilities, suggesting tutorials to improve self-assessment. Ma et al.\cite{ma_are_2024} propose three mechanisms—think, bet, and feedback—for calibrating self-beliefs. Calibrating beliefs about agents is well-studied\cite{ma_who_2023, pinski_ai_2023, louie_novice-ai_2020, kuang_enhancing_2024}. Bansal et al. \cite{bansalUpdatesHumanAITeams2019} note that while updates can enhance AI performance, they may also introduce unexpected behavior, conflicting with users' prior experiences and harming team performance. In addition, team composition influences trust—people trust agents less in the absence of human teammates \cite{schelble2022let}. AI literacy \cite{pinski_ai_2023} and interaction dynamics also matter; trust increases when agents are more likely correct than humans \cite{ma_who_2023} or provide well-timed suggestions \cite{kuang_enhancing_2024}. Errors, their type, timing \cite{erlei_understanding_2024}, and repair strategies \cite{do_err_2023} further affect trust and collaboration. Compared to human trust calibration, less attention is given to how agents calibrate their understanding of humans and themselves. Wang et al. \cite{wang_towards_2021} design agents that monitor both the environment and human actions to refine their understanding. Beyond HCI, approaches like "learning to defer" \cite{madras_predict_2018} allow models to account for human expertise and weaknesses in decision-making. While confidence estimation is widely studied in machine learning, model opacity and complex long-term interactions highlight the need for an HCI perspective.

\textbf{Calibration of Reliance.} Reliance has been a long-standing concern in the HCI community, particularly in promoting appropriate human reliance on agents. Existing studies have shown a close link between reliance and trust. For instance, Ma et al. \cite{ma_who_2023} suggest that inappropriate reliance stems from both misplaced trust in AI and inaccurate self-confidence. However, trust and reliance do not form a simple cause-and-effect relationship. Conceptually, trust reflects cognitive beliefs about agents, whereas reliance manifests behaviorally as the extent to which individuals are influenced by agents, ultimately affecting decision-making outcomes \cite{ma_are_2024}. In complex situations where trust in AI is difficult to establish, participants may also rely on AI without trusting it\cite{qian_take_2024}, such as by appropriating AI resiliently\cite{zhang_resilience_2023} or obeying authority\cite{moran_team_2013}. Therefore, trust calibration is just one factor in achieving appropriate reliance, necessitating broader considerations of human cognition. We organize existing research on reliance calibration from the perspective of the \textit{perception gap} and provide insights into fostering appropriate reliance in HAT by examining three fundamental components of mental models: reasoning, commitments, and beliefs.


For \textbf{false reasoning}, studies examine the influence of explanations, such as fidelity and modality\cite{morrison2024impact}. Linguistic cues also play a role: Zhang et al.\cite{zhang_you_2022} find that assertive expressions increase reliance on AI without boosting trust, while Morrison et al.\cite{morrison2024impact} observe no direct effect of assertiveness on reliance. This highlights the complexity of reliance, influenced by various perceptual factors. Other studies show human factors, like scarcity and time pressure, increase dependence on AI\cite{swaroop_accuracy-time_2024}. For \textbf{mismatched commitments}, Chiang et al.\cite{chiang_are_2023} find that groups rely on agents more than individuals, as some members take on the responsibility of reminding others of the AI’s suggestions. Free-riding also affects commitment\cite{shi_retrolens_2023, gao_collabcoder_2024}. For \textbf{inappropriate beliefs}, both overconfidence and underconfidence in oneself affect reliance on AI\cite{he_knowing_2023, ma_are_2024}, as does belief in the agent’s expertise\cite{zhang_you_2022}. We emphasize that the relationship between trust, reliance, and other factors is complex and cannot be reduced to simple cause and effect. While trust calibration is important, it is not the only path to appropriate reliance. Designing interaction patterns from a cognitive perspective, consistent with Zhang et al.'s\cite{zhang_resilience_2023} view of appropriation, offers a broader design space. In scenarios where trust is difficult to establish, this approach provides a valuable starting point for creating resilient HAT.

\subsubsection{Teamwork of Coordination and Back-up Behaviors}

\begin{figure}
    \centering
    \includegraphics[width=0.75\linewidth]{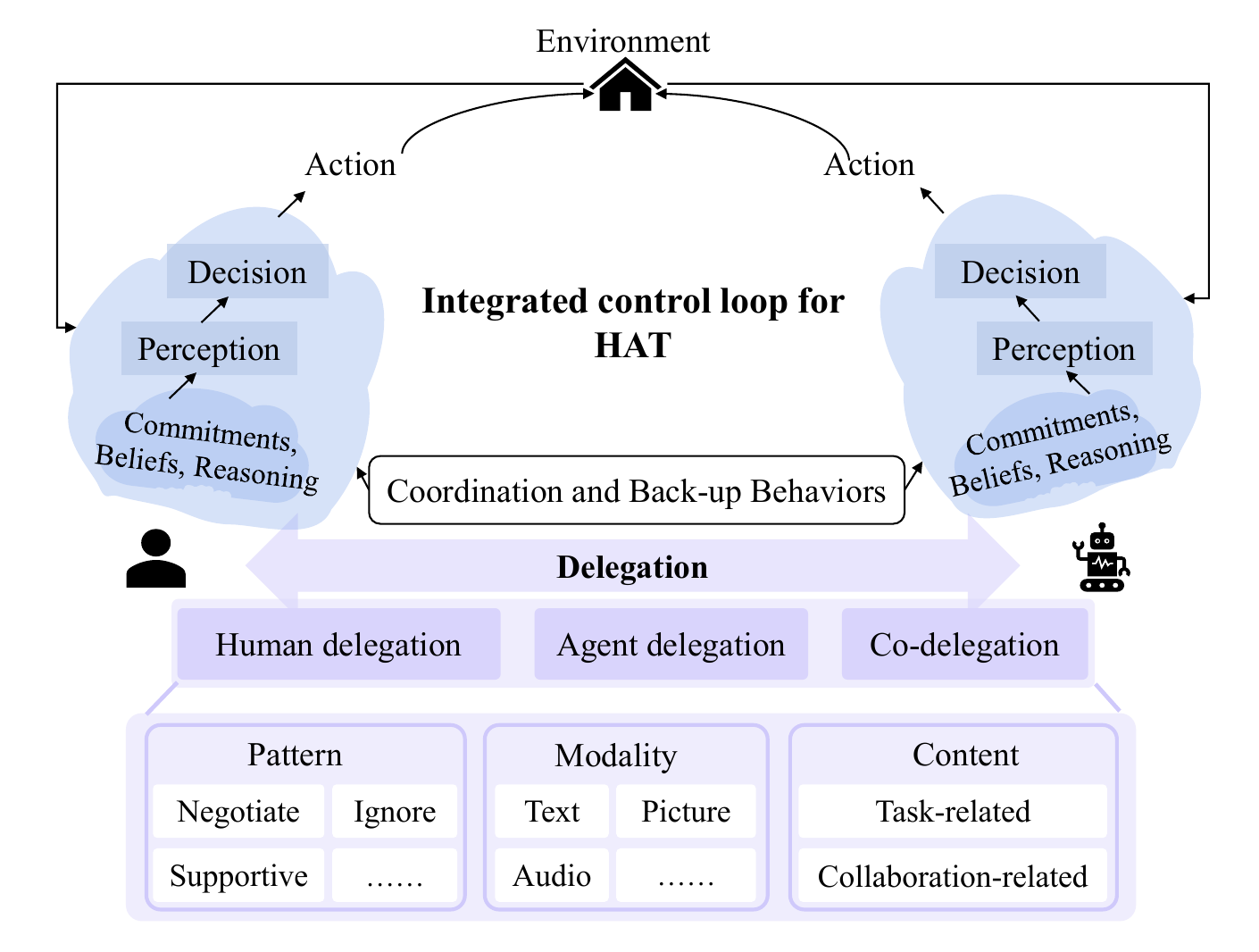}
    \caption{Teamwork of coordination and back-up behaviors. This figure illustrates an integrated control loop for HAT members, showcasing how they perceive, decide, and act based on a delegation mechanism shaped through interactions. Given the variations in interaction patterns, modalities, and content, humans and agents coordinate with each other, forming delegation mechanisms—including human delegation, agent delegation, and co-delegation—to provide back-up behaviors, ultimately enhancing teamwork.}
    \label{fig:enter-interaction}
\end{figure}


For Phase 3 (Team Development), the cognitive aspect requires continuous negotiation to build and refine SMMs, guiding interaction behaviors based on a shared understanding. Once temporarily aligned, these mental models drive coordination and back-up behaviors, reflecting the collaborative efforts of the team. While interactive interfaces (e.g., UI modules, operations) in HATs improve usability, this paper focuses on teamwork aimed at shared goals, particularly regarding control loops for deeper exploration. Through an  control loop for HAT (Fig. \ref{fig:enter-interaction}), team members negotiate task delegation, particularly when one member is unable to complete a task, and other members provide back-up behaviors. Therefore, we focus on delegation mechanisms and also introduce interaction patterns, modalities, and content.


\textbf{Interaction: Pattern, Modality, and Content}. There are some interaction patterns defined, which basically consistent with \textit{the decision and action selections of levels of automation} proposed by Parasuraman et al.\cite{parasuraman_model_2000}, as a combination of one or more of these selections. For example, Sivaraman et al.\cite{sivaraman_ignore_2023} provide interaction patterns from the human perspective, including ignore, negotiate, consider and rely; from the agents' perspective, Cimolino et al.\cite{cimolino2022two} point out that AI can play supportive, delegated, reciprocal and complementary roles in shared control. In terms of interaction modalities, the literature involves text, picture, audio, video, and action modalities, close to half of which is multimodal research. Besides, it involves embodied agents\cite{taylor_coordinating_2019, yuan_social_2022, jung_engaging_2013, lin_it_2020, ye_human---loop_2021, ho_its_2024} and humanoid agents\cite{jung_engaging_2013, louie_novice-ai_2020, suh_ai_2021}. Studies that include audio are mostly related to music besides speech agents, e.g., music composition\cite{martin_intelligent_2016, louie_novice-ai_2020, suh_ai_2021}, real-time music improvisation\cite{mccormack_silent_2019}, and introduced into an inclusive music ensemble\cite{vear_jess_2024}. Most of the literature contains textual modality, serving as an effective medium for interactions and representation of multimedia content\cite{wang_lave_2024}. The content of interactions can be categorized into task-related and collaboration-related, where the former is also task-specific such as providing code\cite{gao_collabcoder_2024}, and the latter promotes the construction of mental models, e.g., the co-creation of abstract drawing\cite{davis_empirically_2016} where agent helps human anticipate and make sense of itself by echoing and mirroring actions.

\textbf{Delegation Mechanisms}.
In the collaboration process of Phase 3, members' meta-task is to establish a reasonable delegation mechanism through dynamic interaction, coordinating with each other and performing appropriate back-up behaviors. We categorize these mechanisms from the literature into human delegation, agent delegation, co-delegation, and deferred mechanisms when delegation is unsuccessful.

\textbf{Human Delegation} refers to humans assigning tasks to agents, a key pattern in collaborative systems \cite{merritt_kurator_2017, cai_human-centered_2019, louie_novice-ai_2020, jeon_fashionq_2021, lee_human-ai_2021, suh_ai_2021, wang_brilliant_2021, gebreegziabher_patat_2023, gu_augmenting_2023, shi_retrolens_2023, lindvall_rapid_2021, overney_sensemate_2024, singh_figura11y_2024} or scaffoldings\cite{wu_ai_2022, gao_collabcoder_2024, dhillon_shaping_2024}, where such delegation is predefined by system designers who structure tasks in advance. In addition, several studies explored how human characteristics affect delegation behavior\cite{he_knowing_2023, ma_who_2023, pinski_ai_2023, zhang_resilience_2023, erlei_understanding_2024}. Pinski et al. \cite{pinski_ai_2023} find that AI-literate users align delegation with their assessments, while Erlei et al. \cite{erlei_understanding_2024} highlight systematic violations of the choice-independence assumption in delegating to superior AI. Effective delegation requires more than enhancing AI capabilities or user understanding; it must account for human uncertainty and decision biases.

\textbf{Agent Delegation} refers to agents assigning tasks to humans when they are unable to complete them or lack confidence, raising two key questions: (1) How do agents decide what to delegate? (2) How does delegation impact humans and the team? \cite{lai_human-ai_2022, moran_team_2013, hemmer_human-ai_2023}. For the first, Bansal et al. \cite{bansalAccuracyRoleMental2019} show that revealing AI error boundaries—particularly its parsimony and stochasticity—helps users form accurate mental models, improving team performance. Lai et al. \cite{lai_human-ai_2022} propose conditional delegation, where humans and agents define trustworthy regions, beyond which tasks are delegated. This interactive negotiation helps establish both agents' capability boundaries and human perceptions of them, suggesting future work on refining these boundaries. The second problem arises because humans retain dominance in the team. Thus, factors beyond performance, such as people's perception of themselves and the nature of work, need more attention\cite{hemmer_human-ai_2023, moran_team_2013}. Hemmer et al.\cite{hemmer_human-ai_2023} observe that regardless of humans' awareness of agent delegation, this behavior improves task performance and satisfaction, with the latter being key to long-term organizational success and self-efficacy. This implies that agent delegation must consider collaboration-related factors such as humans' trust, satisfaction, workload, requiring a deeper understanding of human mental models. The foundation of rational delegation remains the development of SMMs.


\textbf{Co-delegation} refers to real teamwork, where members play reciprocal and even complementary roles. In this scenario, agents may delegate to humans, and humans may also delegate to agents, creating a dynamic combination of human delegation and agent delegation. There are several studies in the review including this mechanism implicitly\cite{zheng_competent_2023, davis_empirically_2016, karimi_creative_2020, yuan_wordcraft_2022, zhou_understanding_2024}, most of which are scenarios of co-creation. A more explicit exploration is the capability-aware shared mental model proposed by He et al.\cite{he_interaction_2023}, which could be further mapped to task assignment, mediating complementary collaboration with relatively few iterations and significantly improving team performance, yet still generally oriented towards agent delegation. We believe that this dynamic, active, and proactive delegation mechanism is a valuable topic for deeper exploration, which can better harness the initiative of both humans and agents to foster complementary collaboration, especially as LLM-enabled agents have a stronger ability to utilize natural language for negotiation.

\textbf{Deferred Mechanisms for Task Delegation.} Before actual delegation, there might be deferred mechanisms. There has been attention in the ML community, e.g., rejection learning is an optional solution that allows models to refuse to make predictions when they are not confidently accurate\cite{cortes_learning_2016}. Madras et al.\cite{madras_predict_2018} further consider human expertise and weakness, proposing learning to defer, which enables models to adaptively judge whether they can accept a delegation or not. In the HCI community, this deferred mechanism embodies more interactivity. Lemmer et al.\cite{lemmer_human-centered_2023} propose human-centered deferred interference, allowing agents to defer and request additional information when uncertain, such as generating follow-up questions, using natural language to revise plans, and asking for a rephrase. This deferred mechanism prompts a shift from traditional one-way delegation to more dynamic co-delegation, reflecting agents more interactively and responsively participating in the team. We believe that this delegation mechanism established by interactive negotiation in the control loop of  HAT, should be an inevitable trend of efficient collaboration between more autonomous agents and humans in the future.

\begin{table}[htbp]
  \caption{The evaluation of HAT across the four development phases}
  \renewcommand{\arraystretch}{1.06}
  \footnotesize
  \label{tab:evaluation}
  \centering
  \begin{tabular}{|m{1.5cm}<{\centering}|m{2.4cm}<{\centering}|m{3.3cm}<{\centering}|m{6.5cm}<{\centering}|}
    \toprule
    \textbf{Phases} & \textbf{Core Indicates} & \textbf{Related indicates} & \textbf{Meanings} \\
    \midrule
    \multirow{2}[0]{*}{\textbf{\shortstack{ \\ \\  Team  \\  Formation} }} 
    & shared goals &  /& referring to shared understanding of  missions, values and the overall goal\\
    & team acceptance & human-like perception\cite{liang_implicit_2019}, partnership\cite{ismail2023public} & referring to the association members perceive between themselves, the team and teammates,and especially humans' perspectives towards agents   \\
 \hline
    \multirow{2}[0]{*}{\textbf{\shortstack{ \\  \\ \\  Task and Role  \\ Development }}}
    & role adherence  & the ability of adhere to the benchmark set\cite{wang2023rolellm}, reliance on agents\cite{gao_collabcoder_2024} & referring to  whether members recognize and adhere to their roles, to ensure clarity of responsibilities\\
    & role competence & self-efficacy\cite{kuttal_trade-offs_2021, hemmer_human-ai_2023, muijlwijk_benefits_2024}, competence/capacity\cite{taylor_coordinating_2019}, usefulness \cite{kim_moderator_2021}, usability\cite{lin_it_2020}  & referring to whether members can fulfill their roles, to identify ability gaps
    \\
  \hline
    \multirow{2}[0]{*}{\textbf{\shortstack{ \\  \\  \\ \\    Team  \\ Development}}}
    & team trust & trust\cite{moran_team_2013}, trustworthiness\cite{kim_engaged_2024}, acceptance\cite{rastogi_deciding_2022}, preference\cite{cai_human-centered_2019}, satisfaction or pride towards agents\cite{park_retrospector_2023}, authority\cite{moran_team_2013},   & referring to the beliefs members hold towards teammates, and especially humans' perspectives on agents, to promote responsibility shifting among members and significantly enhance team cohesion  \\
    & coordination & communication effort and quality\cite{shi_agent_2013,jung_engaging_2013}, communication enthusiasm \cite{rajashekar_human-algorithmic_2024, park_retrospector_2023}, the timing of communication\cite{moran_team_2013, chiang_are_2023}  & referring to the time, frequency, attitude, etc. of team interaction, to reduce misunderstandings, foster consensus, enhance complementarity, and ultimately improve team efficacy  \\
\hline
    \multirow{2}[0]{*}{\textbf{\shortstack{  \\ \\ \\ \\ Team   \\  Improvement}}}
    & team viability & relationship \cite{sadeghian_artificial_2022}, rapport\cite{shamekhi_face_2018, ashktorab_effects_2021}, sense of collaboration\cite{louie_novice-ai_2020}, human satisfaction\cite{sadeghian_artificial_2022}, AI usage continuance intention \cite{cai_human-centered_2019,lee_human-ai_2021,pinski_ai_2023,gu_augmenting_2023} & referring to the rapport between members and their commitment to future collaboration, to promote long-term persistence of the team \\
    & team adaption & /  & referring to the capability of the team to adapt to diverse and dynamic scenarios and tasks  \\
   \\
    \bottomrule
  \end{tabular}
\end{table}

\section{Developmental Evaluation}
\label{Section5}

This section reviews the evaluation of HAT across the four development phases and present some representative metrics for reference (Table \ref{tab:evaluation}). The evaluation in Phase 1 focuses on the initialization of team identity, specifically whether members are aware of the existence of HAT. The evaluation in Phase 2 focuses on individual self-efficacy, namely whether members can adhere to and be competent in their roles. The evaluation in Phase 3 focuses on team efficacy, assessing whether the team achieves good teamwork. The evaluation in Phase 4 focuses on the future improvement of the team, particularly in terms of viability and adaptation.

\subsection{Team Formation}

Based on the definition of HAT (Section \ref{Section1}) and existing literature, two core indicators assess the presence of HAT: shared goals and team acceptance. A shared goal drives execution by aligning members towards a common objective. Team acceptance reflects how well humans accept the team and especially agent members, serving as a critical foundation for cohesive and efficient teams. The \textbf{shared goal} requires team members to reach a consensus on missions, values, and the overall objective, but current research lacks sufficient evaluation in this aspect.  While experimental studies often focus on clear overall goals, aligning missions and values between humans and agents remains a domain-specific challenge that cannot be fully addressed in a single study. Regarding \textbf{team acceptance}, studies \cite{sadeghian_artificial_2022, zheng_competent_2023} highlight that unequal treatment of agents often relegates them to subordinate roles within teams, hindering true team formation. Some studies\cite{wang_brilliant_2021,hayashi_can_2017} make humans negotiate and interact with agents based on \textit{team acceptance}, jointly achieving shared goals. Using the Hanabi\cite{liang_implicit_2019} game, researchers ask human players whether they consider their agent teammates more \textit{human-like}, revealing how team acceptance can enhance teamwork. Similarly, Ismail et al. \cite{ismail2023public} highlight how agents can improve \textit{engagement} and \textit{outcomes} in maternal and child health, showcasing how humans and agents, as partners, work together to achieve shared objectives.

\subsection{Task and Role Development}

In this phase, individual member evaluation is crucial, and assigning specific roles for problem-solving is fundamental to cooperation \cite{mohammadzadeh_studying_2024}, assessed through role adherence and role competence. Role adherence refers to whether members understand and follow their assigned roles, while role competence evaluates their ability to perform within those roles. Regarding \textbf{role adherence}, from the agents' perspective, it is largely technical, such as whether an LLM follows the defined benchmarks for its role \cite{wang2023rolellm}. From the human perspective, adherence can be indirectly reflected by \textit{reliance} on agents, for instance, excessive dependence on agents for convenience \cite{gao_collabcoder_2024} could be seen as a violation of their supervisory role. However, overall, the evaluation of role adherence in HAT members is lacking in the HCI community. As for \textbf{role competence}, it is often measured by the ability of members to accomplish tasks, with task outcomes such as \textit{accuracy}, \textit{efficiency}, and \textit{quality} being very straightforward. Additionally, subjective human feelings are frequently assessed. These subjective metrics include self-awareness of one's abilities through \textit{self-efficacy} \cite{kuttal_trade-offs_2021, hemmer_human-ai_2023, muijlwijk_benefits_2024}, as well as perceptions of the agents' abilities, measured through metrics like \textit{competence/capacity} \cite{taylor_coordinating_2019}, \textit{usefulness} \cite{kim_moderator_2021}, \textit{usability} \cite{lin_it_2020}, and others. However, in later phases of HAT—such as coordination (Phase 3) and adaptation (Phase 4)—roles, responsibilities, and their assignments may change. Some studies intentionally form loosely structured teams without clearly defined roles \cite{kim_moderator_2021, park_retrospector_2023}, raising the question of how to evaluate role adherence and competence in such contexts.

\subsection{Team Development}
To evaluate successful teamwork, we summarize and identify two key indicators: team trust and coordination. Team trust is fundamental to the construction of teams' common ground, which not only promotes resource sharing and responsibility shifting among members but also significantly enhances team cohesion. Team coordination, meanwhile, is a bridge for information transmission, resource allocation, and task negotiation. It can reduce misunderstandings, promote consensus, enhance complementarity, and ultimately improve team efficacy.

\textbf{Team trust} is typically measured by whether humans trust agents, either using direct indicators such as \textit{trust} \cite{moran_team_2013} and \textit{trustworthiness} \cite{kim_engaged_2024} or reflecting human trust through \textit{acceptance} \cite{rastogi_deciding_2022}, \textit{preference} \cite{cai_human-centered_2019}, \textit{satisfaction} or \textit{pride} \cite{park_retrospector_2023} towards agents, as well as \textit{authority} \cite{moran_team_2013} demonstrated by the agents. The evaluation of direct trust indicators is well-established, with scales like Mayer's dimensions of trust \cite{mayer1995integrative} being applied to medical collaboration systems \cite{cai_human-centered_2019}. However, the evaluation of various indirect indicators reflecting trust is more ambiguous. On one hand, these indicators serve specific research needs and only reflect trust to a certain degree. For example, Rastogi et al. \cite{rastogi_deciding_2022} use \textit{agreement with AI} to reflect cognitive bias in AI-assisted decision-making. On the other hand, reliable, validated, and widely accepted scales for these indicators are lacking. For instance, Lin et al. \cite{lin_it_2020} and Park et al. \cite{park_retrospector_2023} used different methods to measure \textit{satisfaction}, which can lead to inconsistencies in understanding. These indicators mainly assess human perspectives on agents, leaving a gap in evaluating true mutual trust within teams. Given that trust calibration involves aligning beliefs with actual abilities \cite{ma_who_2023}, agents' trust in humans also deserves attention. Agents must align their beliefs about human capabilities, and while humans retain ultimate authority, a certain level of doubt about humans may actually be beneficial. Thus, assessing and calibrating agents' trust in humans is a promising area for future research.

\textbf{Team coordination} is usually objectively reflected through the evaluation of communication. Some studies \cite{shi_agent_2013, rajashekar_human-algorithmic_2024} use metrics such as \textit{number, frequency, types, and content of interactions} to form a general impression of coordination. For example, Shi et al. \cite{shi_agent_2013} used these metrics as indicators of communication effort to reveal the positive value of agent metaphors in machine translation-mediated communication, while Rajashekar et al. \cite{rajashekar_human-algorithmic_2024} and \cite{park_retrospector_2023} used them to assess whether humans actively interact and negotiate with agents. Other studies \cite{moran_team_2013, chiang_are_2023} further focus on the timing of communication, evaluating whether perspectives are adequately shared through metrics of \textit{delaying or withholding perspectives}. Regarding the perspectives themselves, the accuracy and clarity of information transmission are key to ensuring smooth communication, though these aspects are difficult to assess and lack quantitative indicators. As for the assessment of\textit{ team efficacy}, which reflects the improvement in coordination, it is more straightforward, typically related to task performance, and includes evaluations of \textit{effectiveness} \cite{park_retrospector_2023}, \textit{quality} \cite{wu_ai_2022}, \textit{efficiency} \cite{lin_it_2020}, and so on.

\subsection{Team Improvement}

Team improvement evaluates the long-term survival and growth potential of the team, which helps ensure that the team maintains competitiveness and adaptability in complex and dynamic environments, demonstrating the team’s vision. Regarding \textbf{team viability}, the rapport between members and their commitment to future collaboration are key indicators of a team’s long-term success. Metrics like \textit{relationship}, \textit{rapport}, and \textit{sense of collaboration} are used to measure members' teaming experience\cite{shamekhi_face_2018,louie_novice-ai_2020,ashktorab_effects_2021,sadeghian_artificial_2022,yuan_wordcraft_2022}, potentially showing future possibilities working together. In particular, human \textit{satisfaction} is regarded as a critical metric for a team's long-term success\cite{sadeghian_artificial_2022}. In addition, metrics like \textit{AI usage continuance intention} and \textit{future use} are directly used as symbols of team viability\cite{cai_human-centered_2019,lee_human-ai_2021,pinski_ai_2023,gu_augmenting_2023}. However, current studies mainly use these metrics to validate AI system usability, without further considering the long-term success of HAT. Regarding \textbf{team adaption}, studies conducted under controlled experimental conditions face challenges in evaluating this aspect. While some research has explored HAT in real-world settings \cite{wang_brilliant_2021, wang_towards_2021, sivaraman_ignore_2023, cai_pandalens_2024}, there remains a lack of assessment on whether these teams truly demonstrate adaptability across diverse environments.

\section{Application of HAT}
\label{Section6}

Given the diversity of HAT tasks, this section divides the scenarios into real-world settings and experimental settings (Table. \ref{tab:task-table1}). \textbf{Real-world settings} highlight HAT's practical applications in daily life, emphasizing agent usability, while \textbf{experimental settings} focus on HAT's fundamental principles, offering tasks and scenarios for future research.
\begin{table}[htbp]
  \caption{Tasks in different real-world and experimental settings}
  \renewcommand{\arraystretch}{1.06}
  \footnotesize
  \label{tab:task-table1}
  \centering
  \begin{tabular}{|m{1.5cm}<{\centering}|m{2cm}<{\centering}|m{2cm}<{\centering}|m{8cm}<{\centering}|}
    \toprule
    \textbf{Setting} & \textbf{Higher-order Task Types} & \textbf{Detailed Task Types} & \textbf{Exemplary Tasks with Reference Paper} \\
    \midrule
    \multirow{17}[0]{*}{\textbf{\shortstack{ \\   \\   \\   \\    \\   \\  \\ \\   \\    \\   \\  \\  \\   \\  \\    \\   \\    \\   \\  \\ \\   \\    \\   \\  \\ Real-world \\ Setting} }} & Personal helper & Daily trivial task & VR shopping \cite{he_towards_2024}; Meeting organization \cite{sadeghian_artificial_2022}; Cross-domain task assistance \cite{sun_intelligent_2016}; Administrative support \cite{salikutluk_evaluation_2024} \\
    & & Companion & Personal curation \cite{merritt_kurator_2017}; Reading \cite{ho_its_2024}; Travel companion \cite{cai_pandalens_2024} \\
    \cmidrule{2-4}
    & Healthcare & Clinical support & Decision support \cite{rajashekar_human-algorithmic_2024,taylor_coordinating_2019,sonntag_towards_2011,cai_human-centered_2019,calisto_assertiveness-based_2023,sivaraman_ignore_2023,yang_harnessing_2023,yildirim_multimodal_2024}; Rehabilitation \cite{lee_human-ai_2021}; Public health \cite{ismail2023public}; Pathology image navigation \cite{gu_augmenting_2023,lindvall_rapid_2021} \\
    & & Care support & Care robot \cite{yuan_social_2022} \\
    \cmidrule{2-4}
    & Vehicle & Automated driving & Non-driving-related activities \cite{berger_designing_2023}; Teleoperation of autonomous vehicles \cite{trabelsi_advice_2023} \\
    & & Aviation & Diversions in aviation \cite{zhang_resilience_2023} \\
    \cmidrule{2-4}
    & Recreational activity & Sport & Marathon running coaches \cite{muijlwijk_benefits_2024} \\
    & & Game & Digital games \cite{cimolino_role_2021,flathmann2024empirically,zhang_investigating_nodate,zhang2021ideal}; Chess \cite{das_leveraging_2020} \\
    \cmidrule{2-4}
    & Research and development & Scientific research & Multi-step retrosynthetic route planning \cite{shi_retrolens_2023} \\
    & & System developing & Programming \cite{ross_programmers_2023,weisz_perfection_2021,gao_coaicoder_2024,kuttal_trade-offs_2021,ruoff_onyx_2023,qian_take_2024}; Usability test \cite{kuang_enhancing_2024} \\
    & & Qualitative coding & Qualitative coding \cite{gebreegziabher_patat_2023,overney_sensemate_2024,gao_collabcoder_2024} \\
    \cmidrule{2-4}
    & Co-creation & Writing & Research questions composing \cite{liu_coquest_2024}; Creative writing \cite{singh_where_2023,qin_charactermeet_2024,clark_creative_2018}; General writing \cite{wu_ai_2022,yuan_wordcraft_2022,dhillon2024shaping}; Screenplays and theatre scripts writing \cite{mirowski_co-writing_2023}; Cartoon-caption writing \cite{kariyawasam_appropriate_2024}; Alternative text descriptions \cite{singh_figura11y_2024} \\
    & & Slides & Slides editing \cite{arakawa_catalyst_2023} \\
    & & Drawing & Creative ideation through AI errors \cite{liu2024smart}; Sketch \cite{lin_it_2020,karimi_creative_2020,williford_recognizing_2020}; Abstract drawing \cite{davis_empirically_2016}; Design \cite{jeon_fashionq_2021,han_when_2024,zhou_understanding_2024,wang_roomdreaming_2024} \\
    & & Music & Music making \cite{vear_jess_2024,martin_intelligent_2016,louie_novice-ai_2020} \\
    & & Improvisation & Improvisational theatre \cite{hodhod_closing_2016}; Movement improvisation \cite{trajkova_exploring_2024}; Music improvisation \cite{mccormack_silent_2019} \\
    & & Video & Video editing \cite{wang_lave_2024}; Short-form video creation \cite{wang_reelframer_2024,kim_unlocking_2024}  \\
 \hline
    \multirow{8}[0]{*}{\textbf{\shortstack{  \\  \\  \\  \\  \\   \\    \\   \\  \\ \\   \\    \\   \\  \\   \\   \\    \\  \\  \\   \\   \\   \\  Experimental \\  Setting}}}
    & Machine learning task& Identification & Non-trivial blood vessel labeling task\cite{bossen_batman_2023}; Image labeling \cite{he_interaction_2023}; Video anonymization \cite{xu_comparing_2023}; Object shape identification task \cite{zhang_you_2022}; Language-based image cropping task \cite{lemmer_human-centered_2023} \\
          &   & Classification & Image classification \cite{morrison2024impact,pinski_ai_2023,hemmer_human-ai_2023,mohammadzadeh_studying_2024}; Biomedical time-series classification \cite{schaekermann_ambiguity-aware_2020}; Classification of heart sound recordings \cite{callaghan_mechanicalheart_2018}; Content moderation \cite{lai_human-ai_2022,jahanbakhsh_exploring_2023}; Deceptive hotel review classification \cite{schemmer_appropriate_2023} \\
        &   & Prediction & Performance prediction task \cite{rastogi_deciding_2022}; Recidivism risk prediction task \cite{chiang_enhancing_2024,chiang_are_2023}; Income prediction \cite{ma_are_2024,ma_who_2023}; CDS risk prediction \cite{rajashekar_human-algorithmic_2024}; Acute MI risk estimation \cite{panigutti_understanding_2022}; Pose estimation \cite{ye_human---loop_2021}; Criminal sentencing estimation \cite{kahr_trust_2024} \\
        \cmidrule{2-4}
            & Resource search and allocation & / & Resource allocation task \cite{schelble2022let,schelble2022see}; Urban search and retrieval (USAR) game \cite{jung_engaging_2013}; Cargo \cite{moran_team_2013}; Debris collection problem \cite{ulusan_rather_2022} \\
            \cmidrule{2-4}
            & Cognitive challenges and intellectual games & Reasoning and logic task & Spatial reasoning and count estimation tasks \cite{cao2023time}; Evaluation of the logical validity of socially divisive statements \cite{danry_dont_2023}; Logical reasoning task \cite{he_knowing_2023}; Logic puzzles \cite{swaroop_accuracy-time_2024} \\
            & & Intellectual game & Hanabi \cite{liang_implicit_2019}; Word guessing \cite{gero_mental_2020,ashktorab_effects_2021}; Quizbowl \cite{feng_what_2019} \\
            \cmidrule{2-4}
            & Discussion & Purpose-oriented & Group ideation (brainwriting) \cite{shaer_ai-augmented_2024,hwang_ideabot_2021}; Group decision making \cite{do_err_2023,kim_engaged_2024,zheng_competent_2023}; Discussion improving \cite{zhang_deliberating_nodate} \\
            & & Context-oriented & Online discussion \cite{kim_moderator_2021}; Embodied discussion \cite{shamekhi_face_2018}; Multilingual communication \cite{shi_agent_2013} \\
            
    \bottomrule
  \end{tabular}
\end{table}

\subsection{Real-world Setting}

The applications of HAT in the real world are extensive, including personal helper, vehicle, healthcare, recreational activity, research and development, co-creation, and so on. \textit{Personal helpers} and \textit{healthcare} are the most common scenarios. For \textit{personal helpers}, agents can facilitate daily tasks like meeting organization\cite{gu_augmenting_2023}, VR shopping\cite{he_towards_2024}, acting as implementers and assist with a task. When personal helpers act as companions, shifting from passive execution to proactive interaction, their social attributes become more prominent \cite{cai_pandalens_2024}. These agents balance task and social capabilities, adopting forms like VR \cite{he_towards_2024}, robots \cite{ho_its_2024}, and head-mounted displays \cite{cai_pandalens_2024}, making them more akin to HAT teammates and promising for daily applications. Specially, \textit{healthcare} demands specialized support in narrow domains, suggesting the need for multiple agents with distinct capabilities rather than a single all-purpose agent. In healthcare, agents are widely used for clinical support, especially decision support\cite{rajashekar_human-algorithmic_2024, taylor_coordinating_2019, sonntag_towards_2011, cai_human-centered_2019, calisto_assertiveness-based_2023, sivaraman_ignore_2023, yang_harnessing_2023, yildirim_multimodal_2024}, and thus focus on the accountability and safety of decision-making. In addition, agents collaborate with care workers\cite{yuan_social_2022}, fostering emotional connections, which places a higher demand on their social capabilities. The \textit{vehicle} domain includes automated driving\cite{trabelsi_advice_2023}, non-driving activities\cite{berger_designing_2023}, and aviation\cite{zhang_resilience_2023}. In \textit{recreational activities}, agents assist coaches in marathon training\cite{muijlwijk_benefits_2024} and frequently appear in digital games\cite{cimolino_role_2021, flathmann2024empirically, zhang_investigating_nodate, zhang2021ideal}, where studies explore human perceptions of agent teammates and communication strategies. In \textit{research and development}, collaborative programming with agents has gained significant attention\cite{ross_programmers_2023, weisz_perfection_2021, gao_coaicoder_2024, kuttal_trade-offs_2021, ruoff_onyx_2023, qian_take_2024}. Additionally, qualitative coding, a key research task in HCI, has inspired HAT-based coding research\cite{gebreegziabher_patat_2023, overney_sensemate_2024, gao_collabcoder_2024}. Notably, Shi et al.\cite{shi_retrolens_2023} propose a collaborative system for chemists and agents to co-design multi-step retrosynthetic routes. In \textit{co-creation} scenarios—spanning writing, drawing, music, video, slides, and improvisation—agents' generative and creative abilities are essential. Arakawa et al.\cite{arakawa_catalyst_2023} explore how agent-generated content in slide editing re-engages users, while Davis et al.\cite{davis_empirically_2016} examine agents in abstract drawing, prioritizing user engagement over output quality. Music improvisation\cite{mccormack_silent_2019}, though part of the broader music domain, exemplifies real-time co-creation in shaping SMMs. These cases underscore the importance of agent interactivity, task capability, and social capability—key to HAT, as discussed in Section \ref{Section4.2}.

\subsection{Experimental Setting}

Most HAT experiments use manually designed, controlled scenarios, often based on machine learning tasks like classification, to study core principles. For instance, Chiang et al.\cite{chiang_are_2023, chiang_enhancing_2024} examine AI-assisted decision-making and its impact on group risk prediction. Content moderation\cite{jahanbakhsh_exploring_2023, lai_human-ai_2022} is also a key focus. In \textit{resource search and allocation}, studies assess HAT performance in multi-objective problems\cite{schelble_i_nodate, ulusan_rather_2022} and explore team composition\cite{jung_engaging_2013}, agent leadership\cite{moran_team_2013}, and mental models\cite{schelble2022let}. For \textit{cognitive challenges and intellectual games}, cooperative games like Hanabi serve as prime testbeds for studying mental models\cite{liang_implicit_2019, gero_mental_2020, ashktorab_effects_2021}. For \textit{discussion scenarios}, research covers purpose-oriented and context-oriented tasks. Purpose-oriented discussions focus on teaming objectives like group ideation\cite{shaer_ai-augmented_2024, hwang_ideabot_2021}, decision-making\cite{do_err_2023, zheng_competent_2023, kim_engaged_2024}, and improving discussion processes\cite{zhang_deliberating_nodate}. Context-oriented studies explore different formats, including online\cite{kim_moderator_2021}, embodied\cite{shamekhi_face_2018}, and multilingual discussions\cite{shi_agent_2013}. Agents increasingly act as social facilitators\cite{shamekhi_face_2018} and moderators\cite{kim_moderator_2021}, moving beyond simple implementation to proactive collaboration. Given real-world complexity, applying these experimental settings requires careful design and thoughtful adaptation.


\section{Discussion}
\label{Section7}
\begin{figure}
    \centering
    \includegraphics[height=4.9cm, width=12.0cm]{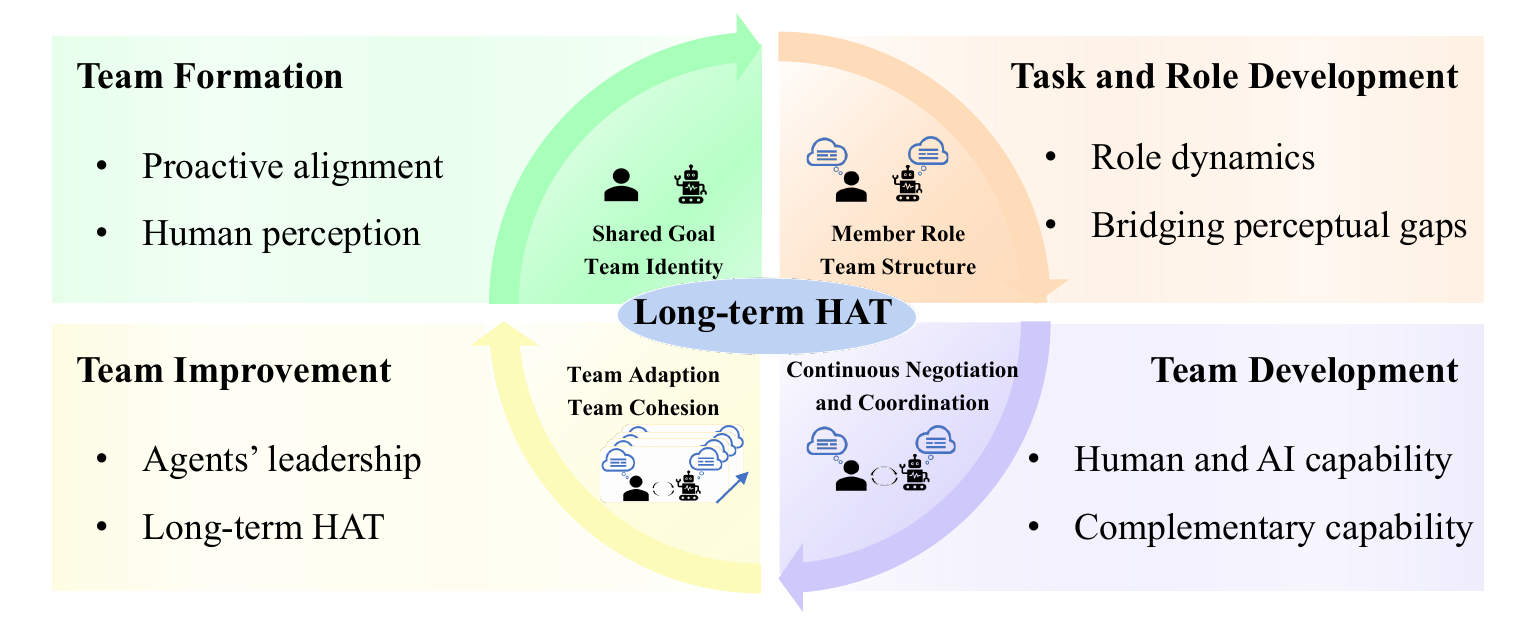}
    \caption{The limitations and future directions of each phase in HAT lifecycle. \colorbox{green!10}{Team Formation}: Existing research often overlooks this phase. Future studies should explore proactive alignment through social interactions and the use of appropriate metaphors to shape human perceptions of agents. \colorbox{orange!10}{Task and Role Development}: Further research is needed to refine team structures, role assignments, and strategies to mitigate humans’ perceptual bias against recognizing agents as teammates—an issue closely linked to the previous phase. \colorbox{blue!10}{Team Development}: While this phase has been extensively studied, future efforts should focus on achieving seamless collaboration between humans and agents. This complementarity should extend beyond task capabilities to include mutual coordination and back-up behaviors. \colorbox{yellow!10}{Team Improvement}: Research in this phase remains limited. Future work should explore the long-term sustainability and adaptability of HAT in real-world environments, as well as the leadership roles that highly autonomous agents might assume, ultimately advancing HAT toward self-management and self-regulation.
    }
    \label{fig:8}
\end{figure}

This section primarily discusses the limitations and future directions of each phase in the HAT lifecycle (Fig.\ref{fig:8}).

\subsection{Team Formation}
\subsubsection{Proactive Alignment and Human Perception of Agents in Team Formation}
In existing research on HAT, the phase of team formation is often overlooked. Agents are typically preconfigured to participate in team tasks, a practice especially common in the design of collaborative systems \cite{gebreegziabher_patat_2023}. For humans, cooperation with agents is usually established through brief instructions, such as written guidelines or verbal briefings before experiments. While aligning task goals in HAT is relatively straightforward, aligning team missions and values differs significantly from human teams. Within the technical community, substantial research has been conducted on AI alignment, covering aspects such as values, interests, and instructions \cite{gabriel2020artificial}. However, this research primarily focuses on broad human alignment rather than alignment within specific team contexts. This gap suggests significant opportunities for further exploration from both technical and HCI perspectives. Specifically, instead of relying on researchers or designers to predefine agents' team missions and values, agents in real-world settings need to engage in proactive alignment through social interaction. With the increasing interactivity of LLM-enabled agents—even their ability to challenge established norms within teams \cite{chiang_enhancing_2024}, such proactive social interaction becomes not only feasible but also essential for effective team formation.

Additionally, during the team formation phase in HAT, unique challenges arise from the nature of agent identities, which do not apply to human teams. One key issue is whether humans will truly perceive agents as team members, thereby forming a genuine team rather than merely using AI tools. Jung et al. \cite{jung2022great} suggest that the non-human metaphorical representations attributed to conversational agents can influence users' engagement, cognitive load, intrinsic motivation, and trust in the agents. Furthermore, Pinski et al. \cite{pinski_ai_2023} demonstrate that humans with AI knowledge can better complement AI, but their willingness to collaborate may decline. This highlights two crucial points: first, individuals may have differing perceptions of agents, and second, the design of agents influences the metaphors associated with them, which in turn affects human cognition. Future research may explore what factors influence whether humans perceive agents as team members. It is important to note that viewing agents as team members does not necessarily mean perceiving them as "human." Hwang et al. \cite{hwang_ideabot_2021} compare the real identity and perceived identity of agents and find that humans contribute more creative self-efficacy when they view their teammates as robots. This suggests that agents' non-human nature can foster psychological safety for human members, enabling them to express themselves more freely \cite{chiang_enhancing_2024}. Therefore, the key to forming HAT may lie in designing agents that are more actively engaged in social interactions, providing appropriate metaphors, and adapting to the specific needs of different contexts, thereby facilitating true further collaboration between humans and agents.

\subsection{Task and Role Development}

\subsubsection{Structural and Role Dynamics in HAT vs. HHT}
HAT exhibits a unique structural configuration compared to traditional human-human teaming (HHT). Although HHT relies on human experience and expertise to dynamically assign roles based on task demands, HAT extends this flexibility by incorporating intelligent agents that can assume specialized roles such as implementers, coordinators, and advisors \cite{zhang_resilience_2023}. This adaptability enhances task performance in complex environments but also introduces challenges such as role confusion and redundancy, particularly in configurations like NvN (multiple humans collaborating with multiple agents) \cite{rajashekar_human-algorithmic_2024}. For instance, in medical teams, if an intelligent agent incorrectly takes over a human’s role, it may lead to role confusion, decision delays and errors, and decreased trust, ultimately reducing collaborative efficiency. In contrast, HHT exhibits a more organic form of role adaptation, where members naturally adjust their roles based on social dynamics, task requirements, and interpersonal relationships \cite{zhang_you_2022}. For example, in emergency response teams, members dynamically shift from executors to coordinators based on situational demands, demonstrating the inherent flexibility of HHT \cite{feng_what_2019}. Although HAT attempts to replicate this adaptability through agent programming, they often lack the tacit situational knowledge that humans employ in role management. Bridging the structural gap between HAT and HHT will require interdisciplinary research that integrates technological advancements with insights from cognitive science. For example, developing reinforcement learning–based dynamic role allocation algorithms could enable agents to adjust their roles in real-time according to changing task demands, thereby reducing role confusion and redundancy. Such approaches would not only enhance team coordination but also help replicate the organic role shifts observed in HHT, paving the way for more resilient and adaptive human-agent collaborations.

\subsubsection{Agent as a Teammate: Bridging Perceptual Bias }
The interaction between humans and agents in HAT is shaped by perceptual differences. Humans often perceive agents as tools or extensions of their capabilities rather than as equal collaborators \cite{kuang_enhancing_2024}. This instrumental view can weaken team cohesion and communication, as agents generally lack the emotional perception and social awareness vital for effective collaboration \cite{wang_lave_2024}. For example, in HHT, non-verbal cues\cite{burgoon2021nonverbal}, empathy, and shared experiences play a crucial role in building trust and resolving conflicts\cite{yang2025socialmind}, whereas these elements are often underdeveloped in HAT. To address these challenges, researchers have emphasized the need to design agents with enhanced communication capabilities, enabling them to understand social cues and contextual nuances \cite{trabelsi_advice_2023}. In creative design teams, for instance, agents that provide real-time feedback and adapt to human preferences can foster greater trust and collaboration \cite{dhillon2024shaping}. Additionally, developing interpersonal mental models—where agents grasp their social roles within the team—can further improve coordination and reduce miscommunication \cite{mathieu2000influence}. To mitigate the perceptual bias of humans, future research should focus on endowing agents with more human-like social behaviors. Enhancements in emotion recognition, empathy modeling, and the ability to process non-verbal cues could help agents better integrate into team settings, thereby reinforcing trust and collaboration. By mimicking key aspects of human social intelligence, intelligent agents can evolve from being perceived merely as tools to becoming genuine collaborators in diverse team environments. Such advancements would be pivotal in achieving truly equitable HAT.

\subsection{Team Development}
\subsubsection{"Complementary Capability": beyond Basic Advantage Sharing}

The complementary strengths of humans and agents enhance team performance \cite{zhang_you_2022}. For \textit{agent capabilities}, Hemmer et al.\cite{hemmer_human-ai_2023} note that models are limited by capacity, data, and unknown outliers, while Wang et al.\cite{wang_reelframer_2024} emphasize generative AI's ability to quickly generate ideas during creative divergence. In terms of \textit{human capabilities}, research highlights traits like humans' ability to make heuristic judgments during creative convergence \cite{wang_reelframer_2024}. These differences are linked to task scenarios, where a trade-off between precision and recall is often seen \cite{xu_comparing_2023}. For instance, Xu et al.\cite{xu_comparing_2023} emphasize a need of high recall in video anonymization, as humans excel at identifying incorrect AI-generated bounding boxes. Additionally, humans have access to contextual information that models often lack \cite{hemmer_human-ai_2023}. For example, Muijlwijk et al.\cite{muijlwijk_benefits_2024} note human coaches' insights into a runner’s psychological profile, which are difficult for models to incorporate. Thus, effective collaboration involves not only sharing information but also task delegation based on the distinct strengths of humans and AI.

The gap between humans and agents with complementary capabilities, and achieving true complementarity in HAT, aligns with the \textit{perception gap} in phase 3. Three factors hinder complementarity in human-delegation scenarios \cite{erlei_understanding_2024}: challenges in enforcing delegation rules, lack of human self- and task-assessment, and violations of task-based choice independence. These issues stem from commitments, beliefs, and reasoning components discussed in Section \ref{Section4.3}. Additionally, Sivaraman et al. \cite{sivaraman_ignore_2023} highlight trust calibration challenges, while Morrison et al. \cite{morrison2024impact} stress the importance of explanations in collaboration. Thus, developing SMM is essential for complementarity, but complete overlap in understanding team functioning and capabilities is neither achievable nor necessary \cite{woehr2003elaborating}. A balance must be struck between constructing SMM and managing the effort involved. To this end, we adapt the concept of backup capability in human-human teams \cite{salas2005there} to the notion of complementary capability in HAT, which involves three processes: \textbf{1) Recognition}: Identifying mismatches between ability and task assignment. \textbf{2) Shifting of work responsibilities}: Transferring tasks to capable members. \textbf{3) Completion}: Ensuring tasks are completed by others. To achieve complementary capability, HAT members' mental models must incorporate perceptions of abilities, responsibilities, and task completion. This concept is linked to the control loop in Fig.\ref{fig:enter-interaction}, aiming for dynamic co-delegation. Enhancing HAT members' complementary capability presents greater challenges for agent design and technology development.

\subsection{Team Improvement}

\subsubsection{Can Agents Get a Higher Position in HAT?}
The level of autonomy is a key attribute for determining whether an agent can be included in a HAT and regarded as a teammate. Current LLM-driven agents already possess high levels of autonomy. Both AutoGPT\cite{significant_gravitas_autogpt_2024} and the Stanford Town experiment\cite{park2023generative} demonstrate the high autonomy potential of LLM agents. So, can agents take on more dominant positions in future HAT? For example, roles are typically at the core of a team, such as leader, expert, or supervisor. Although \textit{the autonomy of an agent is high}, it does not necessarily mean that it can directly enhance its position in the team in all scenarios. The setting of autonomy needs to be flexibly adjusted based on various factors such as task characteristics, user needs, and security laws. In highly efficient, responsive, or hazardous environments, highly autonomous agents may be more favored; In scenarios that require building trust, precise control, or social interaction, it may be more appropriate to reduce autonomy moderately. Secondly, even if an agent is granted high-level permissions, \textit{social and psychological factors may still limit their core role positioning} within the team\cite{walliser2019team,zheng_competent_2023}. Research has shown that although agents can efficiently complete tasks, people still tend to collaborate with humans and view agents as subordinate roles\cite{sadeghian_artificial_2022}. This cognitive bias may stem from inherent beliefs about trust in AI, emotional connections, and traditional role positioning.

\subsubsection{Towards a Long-term HAT}

Existing HAT studies focus on short-term experiments, limiting the understanding of long-term team dynamics \cite{noauthor_development_nodate}. These studies overlook the \textit{duration and importance of a team’s existence}, such as the 36-minute simulation of NeoCITIES teams or the 1-hour DebateBot discussions \cite{kim_moderator_2021}, where short timeframes make self-reported metrics sufficient for assessing SMM and team cohesion. The \textit{development of a team} should be a key indicator for long-term HAT, but few studies address this, making it hard to predict the stability of HAT structures over time. Quantifying the \textit{evolution and persistence} of a team is crucial, and new methods like those proposed by Eloy et al. \cite{eloy2023capturing}, which use near-infrared spectroscopy and recurrence analysis, offer real-time solutions for capturing team dynamics. The research mainly focuses on experimental scenarios and offers design guidelines. Long-term qualitative studies in mature areas like healthcare, autonomous driving, and the military can provide more practical insights for long-term HAT. For instance, Taylor et al. \cite{taylor_coordinating_2019} explore nurses' expectations for robot-assisted decision-making through three months of interviews, aiming to shift power dynamics. In autonomous driving, Chu et al. \cite{chu2023work} highlight the importance of ongoing practice in building trust between safety drivers and autonomous agents. In the military, the GAART system \cite{taberski2021visualizing} uses visualization to improve decision-making and optimize HAT dynamics. All these offer valuable guidance for developing stable, trustworthy, and effective long-term HAT.

\section{Conclusion}
\label{Section8}
In this paper, we present a comprehensive review of the current landscape of HAT research in the HCI community, and propose the T$^4$ framework, a process dynamics framework that integrates both task dynamics and team developmental dynamics. The framework divides the HAT lifecycle into four phases: \textit{team formation}, \textit{task and role development}, \textit{team development}, and \textit{team improvement}. For each phase, we analyze its developmental goals, actions, and evaluation metrics to enhance both task-related and social capabilities, with a particular focus on role allocation and SMM construction in HAT. While significant progress has been made in the \textit{task and role development} and \textit{team development} phases (phases 2 and 3), research on \textit{team formation} and \textit{team improvement} phases (phases 1 and 4) remains limited. We further discuss the limitations and future directions of each phase, emphasizing the need for strategies to strengthen team identity and support long-term adaptability. Ultimately, addressing these gaps will contribute to a more efficient, cohesive, and resilient HAT. Therefore, researchers in the field should adopt a holistic approach to explore the design space and bridge these research gaps. This will not only enhance HATs' self-management and self-regulation capabilities but also improve their adaptability in complex and dynamic real-world environments.



\bibliographystyle{ACM-Reference-Format}

\bibliography{HAT}

\end{document}